\documentclass[%
 reprint,
superscriptaddress,
nofootinbib,
nobibnotes,
 amsmath,amssymb,
 aps, prd
]{revtex4-2}

\usepackage{filecontents}
\usepackage{amsmath}
\usepackage{comment}
\usepackage{multirow}
\usepackage{array}
\usepackage{aas_macros}
\usepackage{graphicx}
\usepackage{dcolumn}
\usepackage{bm}
\usepackage{hyperref}
\usepackage{orcidlink}
\usepackage{cleveref}
\usepackage{comment}
\usepackage{xcolor}

\newcommand{\UCB}{\affiliation{Department of Physics, University of California, Berkeley, Berkeley, CA 94720, USA}}
\newcommand{\UNH}{\affiliation{Department of Physics \& Astronomy, University of New Hampshire, 9 Library Way, Durham NH 03824, USA}}
\newcommand{\WSU}{\affiliation{Department of Physics \& Astronomy, Washington State University, Pullman, Washington 99164, USA}}

\newcommand{\sdetg}{\sqrt{\gamma}}

\graphicspath{{./}{figures/}}

\begin{document}

\preprint{APS/123-QED}

\title{Axisymmetric hydrodynamics in numerical relativity: \\ treating coordinate singularity, artificial heating and modeling MHD instabilities}

\author{Pavan Chawhan~\orcidlink{0000-0002-3694-7138}}\WSU
\author{Matthew D. Duez~\orcidlink{0000-0002-0050-1783}}\WSU
\author{Francois Foucart~\orcidlink{0000-0003-4617-4738}}\UNH
\author{Patrick Chi-Kit \surname{Cheong}~\orcidlink{0000-0003-1449-3363}}\UCB\UNH
\author{Nishad Muhammed~\orcidlink{0000-0001-8574-0523}}\WSU

\date{\today}

\begin{abstract}
Two-dimensional axisymmetric simulations of binary neutron star (BNS) merger remnant are a cheap alternative to 3D simulations.  To maintain realism for secular timescales, simulations must avoid accumulated errors from drifts in conserved quantities and artificial heating, and they must model turbulent transport in a way that remains plausible throughout the evolution. It is also crucial to avoid numerical artifacts due to the polar coordinate axis singularity. Methods that behave well near the axis often break flux-conservative form of the hydrodynamic equations, resulting in significant drifts in conserved quantities.  We present a flux-conservative scheme that maintains smoothness near the axis without sacrificing conservative formulation of the equations or incurring drifts in conserved global quantities.  We compare the numerical performance of different treatments of the hydrodynamic equations when evolving a hypermassive neutron star resembling the remnant of a BNS merger.  These simulations demonstrate that the new scheme combines the axis smoothness of non-conservative methods with the mass and angular momentum conservation of other conservative methods on $\sim 10^2$\,ms timescales of viscous and neutrino-driven evolution.  Because fluid profiles remain smooth in the remnant interior, it is possible to remove artificial heating by evolving the entropy density.  We show how physical heating and cooling terms can be easily calculated from source terms of the conservative evolution variables and demonstrate our implementation.  Finally, we discuss and implement improvements to the effective viscosity scheme to better model the effect of magnetohydrodynamic instabilities as the remnant evolves.
\end{abstract}

\maketitle

\section{\label{sec:level1}Introduction \protect\\}

Binary neutron star (BNS) mergers are multimessenger astrophysical sources emitting gravitational waves and electromagnetic transients\cite{LIGOScientific:2017vwq,LIGOScientific:2020aai,LIGOScientific:2017ync,Baiotti:2017,Burns:2019,Faber:2012,RadiceReview:2020}. Modeling these systems through numerical relativity simulations is crucial for interpreting their gravitational wave and electromagnetic signals. The post-merger gravitational waveform and electromagnetic signals for these systems are sensitive to the fate of the remnant. The remnant can either promptly collapse to a black hole, undergo delayed collapse on a secular timescale, or persist indefinitely as a neutron star. Simulating these systems for long secular timescales is computationally expensive. Although the remnants and their disks are 3D turbulent dynamical systems, their density and velocity profiles on large scales or when time-averaged over multiple dynamical times are expected to approximate axisymmetric dynamical equilibria undergoing slower secular evolution. An attempt to model the remnant using 2D axisymmetric evolutions that resemble 3D evolutions is worth consideration since these simulations significantly reduce computational costs.  With 2D simulations, it is also possible to study the remnant for secular timescales and cover the relevant parameter space more thoroughly with many inexpensive simulations.

It must be acknowledged that even an optimal axisymmetric simulation will omit information and be unable to capture inherently 3D effects. Magnetohydrodynamic (MHD) flows are quantitatively and qualitatively different in 2D vs 3D, as evident by opposite turbulent energy cascades\cite{Batchelor:1969abc,Davidson:2015abc} and the 2D antidynamo theorem\cite{Cowling:1933abc,Moffatt:1978abc}. Even in the absence of a magnetic field, systems such as inclined and precessing NSs could only be studied using 3D simulations. Axisymmetric systems can also be subject to nonaxisymmetric instabilities, and axisymmetic modes are necessarily inaccessible to 2D simulations.

Nevertheless, the feedback of nonaxisymmetric effects on the azimuthal mean fields can be incorporated, at least approximately, using mean-field methods.  The matter fields are then split into an axisymmetric ``background'' and nonaxisymmetric ``turbulence''.  2D simulations only evolve the background fields, but their evolution equations are modified to incorporate feedback from turbulence. Turbulent momentum transport, for example, can be modeled using an effective viscosity. One challenge of long-time 2D simulations is to incorporate these nonaxisymmetric feedbacks in a realistic way.

A realistic axisymmetric simulation must also avoid accumulating significant drifts due to numerical error in extensive conserved quantities like total mass or total angular momentum, since these should not vary with time for an isolated system.  Usually such extensive quantities are given by the integral of an associated density \textit{A} over the entire domain. Here the quantity \textit{A} is given by an evolution equation of the type,
\begin{equation}
    \partial_t A + \partial_x F = S,
\end{equation}
which is referred to as a flux-conservative equation, where $F$ is the flux, and $S$ is the source term, associated with the quantity \textit{A}. In the absence of the source term and of boundary fluxes, these global quantities, like total mass or angular momentum, are conserved. Any drift in mass or angular momentum changes the system under consideration and may significantly affect the dynamics. For example, a small relative change of mass or angular momentum might trigger a delayed collapse or represent a significant fraction of the outflow mass.

A polar coordinate system (spherical-polar or cylindrical-polar) best matches the symmetry of an axisymmetric system.  However, as discussed in section \ref{Conservative_Scheme}, numerically evolving these equations in spherical-polar coordinates using finite-difference often leads to unphysical behavior close to the axis due to the coordinate singularity on the axis. There are several numerical schemes that handle this unphysical behavior, most of which break the flux-conservative form of the above equation and therefore lose accuracy. These non-conservative schemes include some variants of the cartoon method\cite{Alcubierre:2001} and reference-metric formalism\cite{BaumgarteEtAl:2013}, widely used for general relativistic hydrodynamic simulations.  Some early 2D simulations used conservative methods and dealt with the coordinate singularity by adding dissipation near the axis\cite{Shibata:2000,Matt:2004}.

In this paper we introduce a flux-conservative scheme that we call ``modified conservative" that enforce zero flux on the axis and combines the strengths of conservative and non-conservative methods. The basic idea with the later is to maintain conservative form but modify the fluxes close to the coordinate singularity such that the divergence of the new fluxes matches that of a non-conservative method that analytically treats the singularity. We find this scheme to have much better conservation properties while also eliminating axis instabilities. 

Our modified handling of conservative fluxes limits the drift in the conserved baryonic mass and angular momentum observed with  non-conservative methods. We also introduce a technique to mitigate numerical heating and to more realistically model angular momentum transport. We derive and implement an evolution equation for entropy density for use in regions where smooth fluid profiles are expected (i.e. in the absence of non-smooth entropy-generating features such as shocks and current sheets).  We present a general formula for the entropy density source term for any spatially smooth process in terms of the source terms added to the standard hydrodynamic evolution variables. This further improves the accuracy of the simulation by diminishing artificial heating due to numerical viscosity. We also discuss and model angular momentum transport due to turbulence created by MHD instabilities and include it in our simulations. These models are designed to model turbulence for two particular instabilities that are expected to be important for BNS remnants: the magnetorotational instability (MRI) and the Tayler-Spruit instability(TSI).

The paper is organized as follows. In Section \ref{InitialDataAndEvolutionMethods} we outline the methods used to build initial data for an equilibrium neutron star resembling a BNS remnant, and we outline numerical schemes used for evolutions. In Section \ref{DifferentFluxSchemes}, we discuss flux-conservative and non-flux conservative schemes, and our modified conservative scheme. In Section \ref{EntropyEvolution} we derive the evolution equation for entropy density and demonstrate its behavior by implementing it for the remnant. We summarize and conclude in Section \ref{Conclusion}.  In the Appendix, we discuss angular momentum transport due to MHD instabilities. Throughout the paper we use units in which $c = G = M_{\odot} = k_b = 1$, unless stated otherwise.

\section{Initial Data and Evolution Methods} \label{InitialDataAndEvolutionMethods}
\subsection{Evolution Code}
We use the Spectral Einstein Code (SpEC)\cite{SpEC} to evolve Einstein's equations and the general relativistic equations of ideal hydrodynamics. 
The general relativistic hydrodynamic equations are evolved using the conservative variables
\begin{eqnarray}
\label{eq:fluid-density}
\rho_\star &\equiv& - \sdetg n_\mu n^\nu \rho = \rho W \sdetg, \\
\label{eq:fluid-energy}
{\mathbf\tau} &\equiv& \sdetg n_\mu n_\nu T_F^{\mu\nu} - \rho_\star \\
&=& \rho_\star \left(hW - 1 \right) - P\sdetg, \\
\label{eq:fluid-momentum}
S_i &\equiv& - \sdetg n_{\mu} T_{F\ i}^\mu \rho_* h u_i \, \\
\label{eq:rhoye}
(\rho Y_e)_\star &\equiv& \rho_\star Y_e\ ,
\end{eqnarray}
where $n^\mu$ is the normal vector to the spatial slice, $\rho$ the baryonic density, $P$ the gas pressure, $h$ the specific enthalpy, $Y_e$ the reduced electron fraction, $T_F^{\mu\nu}$ the corresponding perfect fluid stress tensor, $W = \sqrt{1 + \gamma^{ij} u_i u_j}$ the Lorentz factor, and $\gamma$ the determinant of the 3-metric $\gamma_{ij}$.

A multidomain grid of points is used to evolve Einstein's equations pseudospectrally in a generalized harmonic formulation \cite{Lindblom:2006}, while the general relativistic hydrodynamics equations are evolved on a finite difference grid. The finite difference grid uses a high-order shock capturing method: a fifth-order WENO reconstruction scheme\cite{Liu:1994,Jiang:1996} to interpolate from cell centers to cell faces and an HLL approximate Riemann solver\cite{HLL:1983} to calculate numerical fluxes at cell faces.

Both the Einstein's equations and the fluid equations are evolved in time using a third order Runge-Kutta algorithm each having the same time step. The source terms are communicated between the two grids at the end of each time step. At the intermediate time steps, a linear extrapolation from the previous steps is done to calculate the source terms to second-order accuracy in time. The interpolation from the pseudospectral grid to the finite difference grid is done by refining the pseudospectral grid by a factor of 3 and doing fifth order interpolation from the refined grid. The interpolation from finite difference to pseudospectral grid is done using a third-order polynomial interpolation.

We use the 2D axisymmetric version of SpEC for the metric and hydrodynamic evolution, each carried out on a 2D grid~\cite{Jerred,Nishad:2024}.  Both 2D grids are on a meridional cut through the presumed axisymmetric system, with the finite difference grid covering one side of the axis and the pseudospectral grid covering both sides symmetrically. We further impose reflection symmetry about the equator on the finite-difference grid and evolve only a quadrant with the field variables in the other quadrant obtained from the evolved quadrant with appropriate symmetry factors.  The pseudospectral grid is divided into subdomains, each with its own set of basis functions and associated colocation points. 

Although each grid is 2D, the tangent space on which vectors live is still 3D, and it is different for spectral and finite difference grids.  In particular, they use a different basis vector pointing out of the plane.  The hydrodynamic equations are written in a polar coordinate system, for which the third coordinate is azimuthal angle $\phi$.  For the pseudospectral spacetime evolution, the third coordinate is the third Cartesian coordinate.  Both grids use the axisymmetry condition, $\mathcal{L}_{\partial/\partial\phi}T=0$ for calculating the perpendicular derivatives. For hydrodynamic grid for which the third coordinate is the global azimuthal angle $\phi$, the axisymmetry condition translates to $\partial_\phi T^{\mu_1 \mu_2 ...} \text{}_{\nu_1 \nu_2 ...} = 0$.
For the pseudospectral grid that uses the cartoon method \cite{Alcubierre:2001,Pretorius:2005,Hilditch:2015}, the out-of-plane derivatives do not necessarily all vanish and can be analytically calculated from the field and their in-plane derivatives on the meridional plane using the symmetry condition.

As in Jesse~\textit{et al.}\cite{Jerred}, the finite difference grid is divided into separate 2D subdomain patches, each with its own local coordinate system, $x^i_L$, related to a global coordinate system, $x^i_G$, by a map that controls the embedding of the domain in the global space. Each patch is a square covering $[-1,1]^2$ in local coordinates.  The evolution is performed in the local coordinate system of each individual patch/subdomain which is then transformed back to the global coordinates for any necessary communication between subdomains. The communication occurs through synchronizing values at the end of each time step and time derivatives at the end of each substep in the ghost zones of each subdomain. We also create and extend ghost zone points beyond any symmetry boundaries to impose boundary and symmetry conditions. We refer readers to Jesse~{\it et al.}\cite{Jerred} for further details. 

We consider linear and polar maps for relating local and global grid points. A linear map ($x^i_G = a_i x^i_L + b_i$) corresponds to grid points that are globally rectangular whereas a polar map corresponds to grid points that are globally wedge of a circle. Since the tangent space on which the vector lives is 3D (azimuthal systems can have azimuthal velocity), the Jacobian map is also 3D with the third component in the local coordinates being the global azimuthal $\phi$. Thus by construction, the local coordinates in which we evolve, $(x,y,\phi)$, will have coordinate singularity near what could be identified as the axis in global coordinates($\varpi = 0$).

As in Jesse~\textit{et al.}\cite{Jerred}, the neutrino fields are evolved using a grey two-moment closure scheme as described in \cite{Foucart:2015vpa,Foucart:2016rxm} for the long-time simulation and for the tests with entropy density evolution. All the simulations in section~\ref{DifferentFluxSchemes} that are done for 5\, ms are carried out without any neutrino physics since matter-neutrino interaction would have negligible effect on the remnant on such short timescales. In Foucart~\textit{et al.}~\cite{FrancoisNeutrinoTransport:2024}, improved methods for the use of two-moment schemes developed by Radice~\textit{et al.}~\cite{RadiceM1:2021} were adapted in SpEC for 3D evolutions. One of these improvements involves changes to the treatment of numerical fluxes in high-opacity regions, referred to here as ``Radice'' fluxes.  We implement these changes to the calculation of numerical fluxes in high-density regions for the 2D axisymmetric version of SpEC for the neutrino evolution.  We find using the previous high-opacity fluxes (described in~\cite{Foucart:2016rxm}) results in an instability, with neutrino quantities rising sharply in a region close to the axis (which then creates artifacts in the fluid temperature), while these fields remain smooth when evolved using Radice fluxes. 

Turbulent angular momentum transport is incorporated via an effective viscosity.  The kinematic viscosity is designed to model MRI turbulence and to behave realistically in both the inner region (which is MRI stable and for which transport is suppressed) and the outer disk-like region. We also consider a model for angular momentum transport due to TSI instability. However, we find the Brunt-V{\"a}is{\"a}l{\"a} frequency of this particular remnant to be smaller than the angular frequency in convectively stable regions, so this remnant is outside the domain of applicability of the well-known scaling relations for TSI angular momentum transport. We thus only include angular momentum transport due to MRI turbulence for this study.  Details for the MRI and TSI turbulent momentum transport models are provided in the Appendix.

\subsection{Initial Data and equation of state}
\label{sec:initial_data}

We model the neutron star using the DD2 equation of state(EoS) \cite{DD2}, which provides pressure $P$ and specific enthalpy $h$ as a function of baryonic density $\rho$, temperature $T$ and reduced electron fraction $Y_e$. It predicts radius $R_{\rm NS}=$13.1\,km and the dimensionless tidal deformability $\Lambda=860$ for a 1.35$M_{\odot}$ neutron star. The initial data is motivated from Vincent \textit{et al.}'s \cite{Vincent:2020} post-merger remnant data and is constructed using the code of Cook, Shapiro and Teukolsky \cite{Cook:1992,Cook:1994}, hereafter  ``RotNS''. RotNS constructs axisymmetric equilibrium configurations with central baryonic density $\rho_c$, 1D equation of state $P(\rho)$ and $h(\rho)$, rotation law $j = j(\Omega)$ or $\Omega = \Omega(j)$, and total angular momentum as inputs. Since it requires a 1D EoS for equilibrium construction, we construct one-dimensional cuts of DD2 by imposing two conditions to determine $Y_e$ and $T$ for each $\rho$. The first condition is beta equilibrium:  $\mu_p(T,\rho,\text{Y}_e) +\mu_e(T,\rho,\text{Y}_e)=\mu_n(T,\rho,\text{Y}_e)+\mu_{\nu}(T,\rho,\text{Y}_e)$, where we take the electron neutrino chemical potential $\mu_{\nu}$ to be zero. For the second condition, we make an explicit choice for specific entropy such that the remnant's central temperature is the similar to one of Vincent \textit{et al.}'s \cite{Vincent:2020} post-merger remnant and set $s(T,\rho,\text{Y}_e) = 1 \text{k}_{B}\text{/baryon}$. Binary neutron star simulations predict merger remnants with a non-monotonic rotation profile, $\Omega(r)$, with a peak some distance away from the rotation axis\cite{Kastaun:2014fna,Hanauske:2016gia,de2020numerical,iosif2021equilibrium,iosif2022models,cassing2024realistic,staykov2023differentially}. Rotation laws that do capture this $\Omega(r)$ profile shape have been constructed by Ury{\={u}}~{\it et al.}~\cite{Uryu:2017obi}. With the 1D EoS and rotation law specified, we construct a hypermassive NS with total mass $M = 3.19 M_{\odot}$, and total angular momentum $J = 5.78$. The equilibrium for this star lies on the stable branch of its constant angular momentum space of equilibria. We refer readers to Muhammed~{\it et al.}~\cite{Nishad:2024} for details.

The spectral evolution grid consists of 87 concentric annuli covering radii 0.44 --29,600\,km with each annuli having 26 angular gridpoints.  The finite difference grid is comprised of square blocks with 120 grid points on each side, which in global coordinates roughly correspond to a square with side length 18\,km.  Both grids are chosen such that no colocation point lies on the axis.  

\section{Different Flux Schemes}\label{DifferentFluxSchemes}
We now explain in detail the different fluid evolution schemes, evaluating their conservation and axis-handling properties.  Then we carry out simulations of the remnant system described above using the different schemes. These simulations are done for 5\,ms, which is sufficient to see the differences between schemes. Note that Sections \ref{Conservative_Scheme},\ref{Factored_Scheme},\ref{ModCons_Scheme} implement finite-differencing when evaluating divergence of the flux. 

\subsection{Conservative Scheme}\label{Conservative_Scheme}

Let us consider a 1D flux-conservative equation for a quantity scalar $A$ with source terms:
\begin{equation}\label{ConservativeFluxForm}
    \partial_t A + \partial_x F = S\ ,
\end{equation}
where $F$ is the associated flux.  The spatially discretized, finite difference form of this equation on a uniform grid with grid spacing $\Delta x$ at gridpoint $i$,
\begin{equation}
    \partial_t A_i = \Delta x^{-1}(F_{i-1/2}-F_{i+1/2}) + S_i\ ,
\end{equation}
is conservative in the following sense.  For evolution variables with $S=0$ ($\rho_\star$ and $S_\phi$, in our 2D simulations) the spatial integral
\begin{equation}
M_A\equiv \sum_i A_i \Delta x
\end{equation}
is conserved up to roundoff error, with deviations only due to fluxes on boundaries and adjustments to evolution variables in the low-density ``atmosphere''.  This is because each interior cell face flux contributes to $\partial_tM_A$ twice with opposite sign:  once as taking from one cell and another as adding to the adjacent cell.  Note $M_{\rho_\star}=M_0$ is the baryonic mass, and $M_{S_\phi}=J$ is the fluid angular momentum.

The flux-conservative form of hydrodynamics evolves variables that are proportional to $\sqrt{\gamma}$ where $\gamma$ is the determinant of the spatial 3-metric. In SpEC, the hydrodynamic evolution variables are evolved in local coordinates (with 3D polar coordinate basis vectors) that are then transformed to global coordinates (with 3D Cartesian coordinate basis vectors) for any necessary communication.  Under global to local transformations, the metric determinant (a scalar density of weight 1) transforms as $\sqrt{\gamma_L} = \mathcal{J} \sqrt{\gamma_G}$, where $\mathcal{J}$ is the determinant of the Jacobian of the local-to-global map.

Due to the coordinate singularity in local coordinates, the Jacobian and therefore the tensor densities in local coordinates vanish on the points that map to the axis in global coordinates and approach zero at nearby points. The Jacobian could be written as $\mathcal{J} = \varpi \tilde{\mathcal{J}}$, where $\varpi$ is the distance from the axis (also a global coordinate) and $\tilde{\mathcal{J}}$ is non-zero on the axis. $A$ can also be factored as $A \sim \varpi \tilde{A}$, where $\tilde{A}$ is non-zero on the axis. Combining this with the flux-conservative equation (\ref{ConservativeFluxForm}), in absence of source terms gives,
\begin{equation}\label{ConsTildeA}
    \partial_t \tilde{A} = -\frac{\partial_x F}{\varpi}.
\end{equation}
For the time derivative of $A$ to be well defined, the partial derivative $\partial F/\partial x$ must go to zero close to the axis. The numerical approximation of $\partial F/\partial x$ will be a sum of its true value and the truncation error, where the truncation error might be some non-zero value close to the axis. Therefore, the right-hand side in Eq.~\ref{ConsTildeA} in the limit, $\varpi \rightarrow 0$ could be some large value, so $\tilde{A}$ can grow over time near the symmetry axis.

\begin{figure}[h!]
    \centering
    \includegraphics[width=1\linewidth]{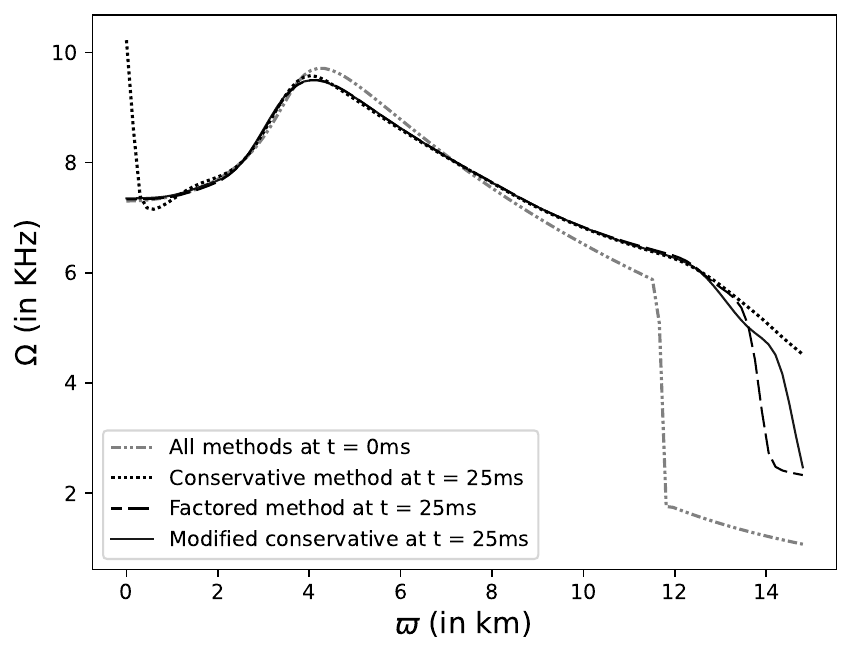}
    \caption{Angular velocity $\Omega$ profiles at different times as a function of cylindrical radius, $\varpi$, for different schemes plotted on the equator.}
    \label{fig:OmegaOnLine}
\end{figure}

Fig.~(\ref{fig:OmegaOnLine}) shows the angular velocity of the differentially rotating star after 25\,ms of evolution on the equatorial line for an evolution using a conservative scheme (dotted line).  We observe large angular velocity after 3.4 central periods in a narrow region close to the axis. Such ``glitches'' were also observed in Jesse~\textit{et al.}~\cite{Jerred}. We remark that not all conservative schemes lead to such glitches. As discussed in Section~\ref{OtherMethods}, conservative schemes discretized using finite-volume methods mitigate axis issues while also maintaining conservation in global quantities. 

\subsection{Factored Scheme}\label{Factored_Scheme}
The axis instability could be treated analytically by factoring out the singular terms in $F$ prior to computing the divergence.  For a component of the flux $F^i$, that contains factors of $\varpi$, an appropriate factoring will be of the form
\begin{equation}\label{FactoredFluxes}
    F^i = \varpi^n \tilde{F^i}\ ,
\end{equation}
where $\tilde{F}$ is the $\varpi$-factored form of the flux, and the integer \textit{n} will depend on the flux and its tensor index $i$~\cite{Jerred}. One can then take the divergence of the factored flux and apply the product rule, which gives

\begin{equation} \label{FactoredFluxForm}
     \partial_i (\varpi^n \tilde{F^i}) = \varpi^n\partial_i \tilde{F^i} + n\varpi^{n-1}\frac{\partial \varpi}{\partial x^i}\tilde{F^i}\ .
\end{equation}
The right-hand side of equation (\ref{FactoredFluxForm}) contains two terms.  The first involves divergence of the factored flux and would involve computing the fluxes at cell-faces. The second term involves the coordinate derivative of $\varpi$ (which is already computed in SpEC as a component of the local-to-global map Jacobian matrix) multiplied by the factored flux, both evaluated at the cell center and obtained analytically.  The evolution equation for $\tilde{A}$ is then
\begin{equation}\label{FacTildeA}
     \partial_t \tilde{A} = -\partial_x \tilde{F} - \frac{n \tilde{F}}{\varpi}\partial_x \varpi
\end{equation}
Unlike Eq.~(\ref{ConsTildeA}), the divergence of $\tilde{F}$ has no $\varpi$ factor in the denominator, whereas the second term involves no finite difference.  Thus, the right-hand side is well-behaved close to the axis.

In Fig.~\ref{fig:OmegaOnLine}, we plot the angular velocity profile on an equatorial plane for the remnant system evolved using different schemes. The angular velocity profile obtained by evolving with a factored scheme develops no sharp feature close to $\varpi = 0$ and is well-behaved.

\begin{figure}[h!]
    \centering
    \includegraphics[width=1\linewidth]{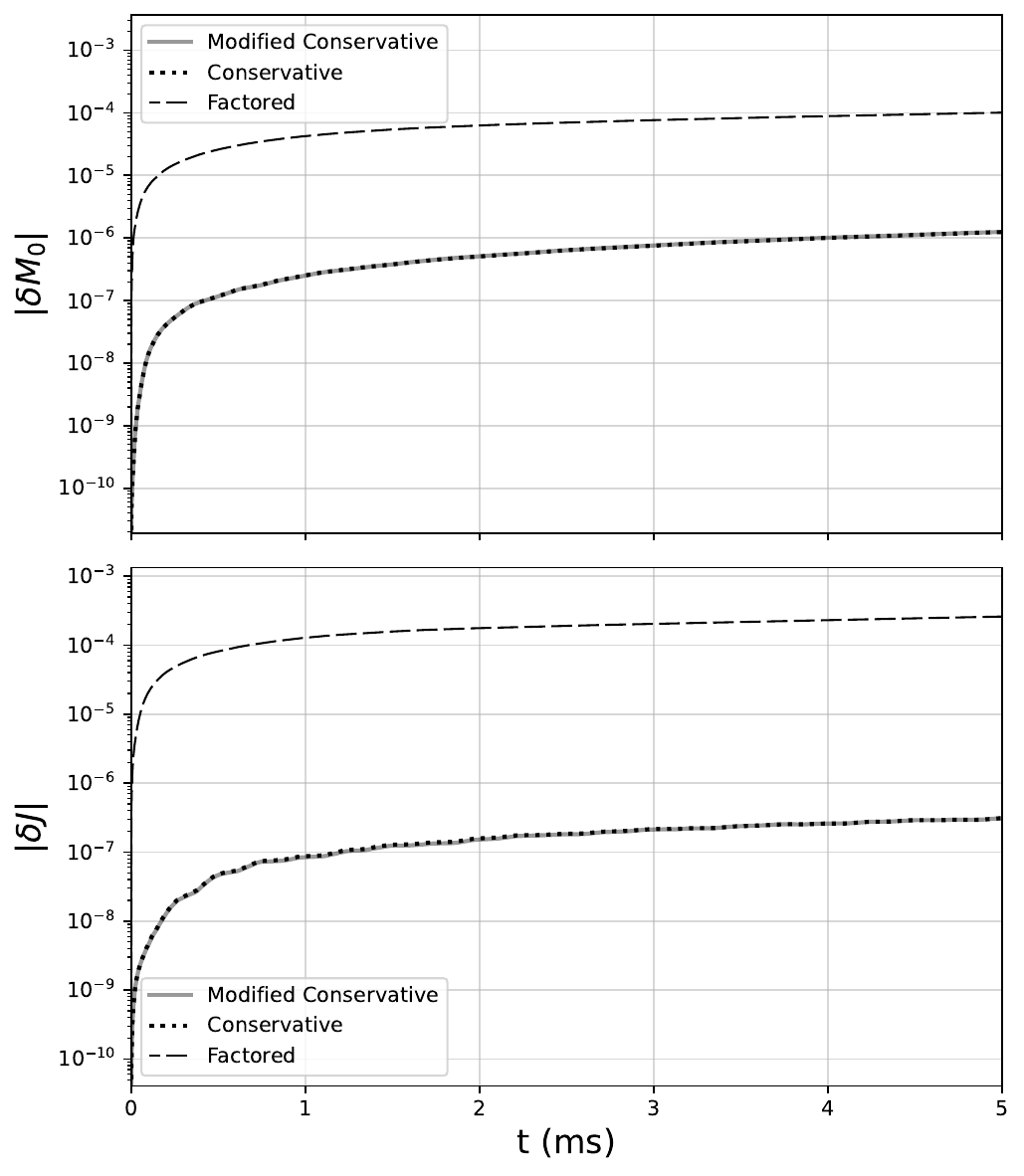}
    \caption{Drift in baryonic mass(\textbf{top}) and angular momentum (\textbf{bottom})for conservative and factored schemes plotted against time. The angular momentum drift is negative in sign for modified conservative and factored schemes.}
    \label{fig:TotalMassAndAngMomError}
\end{figure}

However, the term $n\varpi^{n-1}\frac{\partial \varpi}{\partial x^i}\tilde{F^i}$ in Eq.(\ref{FactoredFluxForm}) appears formally as a source term (cf. Eq. \ref{ConservativeFluxForm}) and so breaks the conservative-flux form of the equation. Therefore, the associated global quantity will not be conserved to roundoff error, but only to truncation error.  This can be seen in Fig~(\ref{fig:TotalMassAndAngMomError}), where we plot the drift $\delta X$ in global quantities $X$ (total mass and total angular momentum) from their initial values at $t =0$, $\delta X = [ X(t) - X(0) ]/X(0)$ .  We notice that the drift in global quantities for the factored scheme is about 3 orders of magnitude bigger than conservative scheme. 

We also test the dependence of the accuracy of conserved quantities on the finite difference grid resolution. For the modified conservative scheme, where one expects conservation up to roundoff error, we find that conservation does not depend on the grid resolution. For the factored scheme, conserved quantities like the total baryonic mass $M_0$, which is the integral of density over the entire domain, are conserved to truncation error and would be expected to converge to a constant with grid resolution.  In axisymmetry, fluid angular momentum $J$ can also be expressed as the integral of the fluid azimuthal momentum density $S_{\phi}$, and $S_{\phi}$ obeys a flux-conservative evolution equation with no source.

This artificial change of $M_0$ and $J$ often manifests as a secular drift, so it accumulates in time.  The drift in global quantity $X$ can be quantified by the relative rate $\Delta X\equiv \langle dX/dt\rangle/X_0$, where $\langle \rangle$ indicates an average over a finite time range, and $X_0$ is the initial value of $X$.

\begin{figure}
    \centering
    \includegraphics[width=1\linewidth]{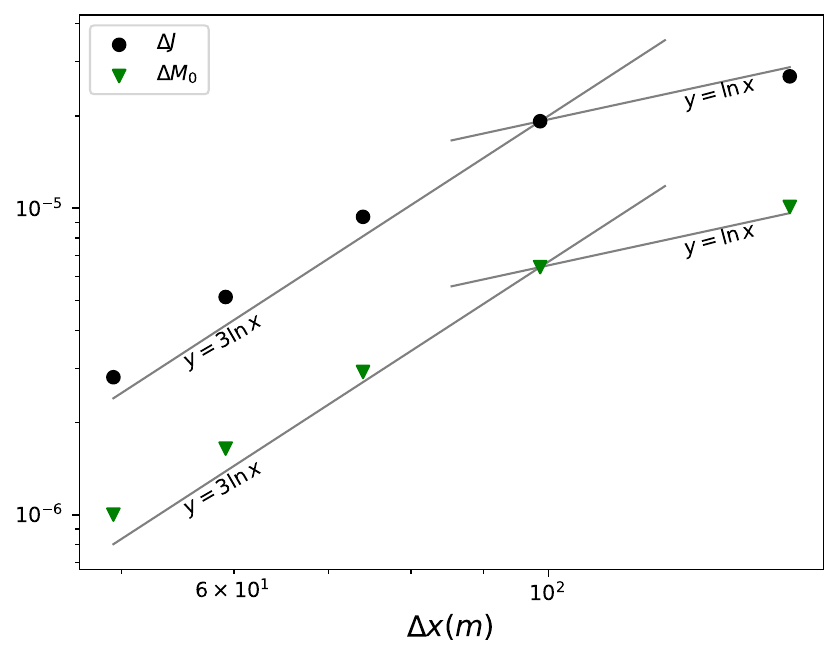}
    \caption{The rate of change of the relative error for total angular momentum (filled circles in black) and total baryonic mass (inverted triangles in green) for 5 different finite difference resolutions, $\Delta x = \{ 148\text{ m},99\text{ m},74\text{ m},59\text{ m},49\text{ m} \}$}.
    \label{fig:ConvergenceTest}
\end{figure}

In Fig.~\ref{fig:ConvergenceTest}, we plot $\Delta M_0$ and $\Delta J$ for 5 different grid resolutions for the factored scheme. The rates are calculated by least-square fitting for the relative drift $[X(t) -X(0) ]/ X_0$, as a function of time. For the test, we vary the resolution of the finite difference grid, a square domain of side length 18\,km covered by $N^2$ grid points, with $N = \{ 120,180,240,300,360\}$. We see roughly third-order convergence for all resolutions except near $\Delta x = 148$\,m.  At low resolutions [$\Delta x \approx 148$m ($N \approx 120$)], convergence is first-order, implying the grid is not in convergent regime. 

\subsection{Modified Conservative Scheme}\label{ModCons_Scheme}
\begin{figure}[h!]
    \centering
    \includegraphics[width=1\linewidth]{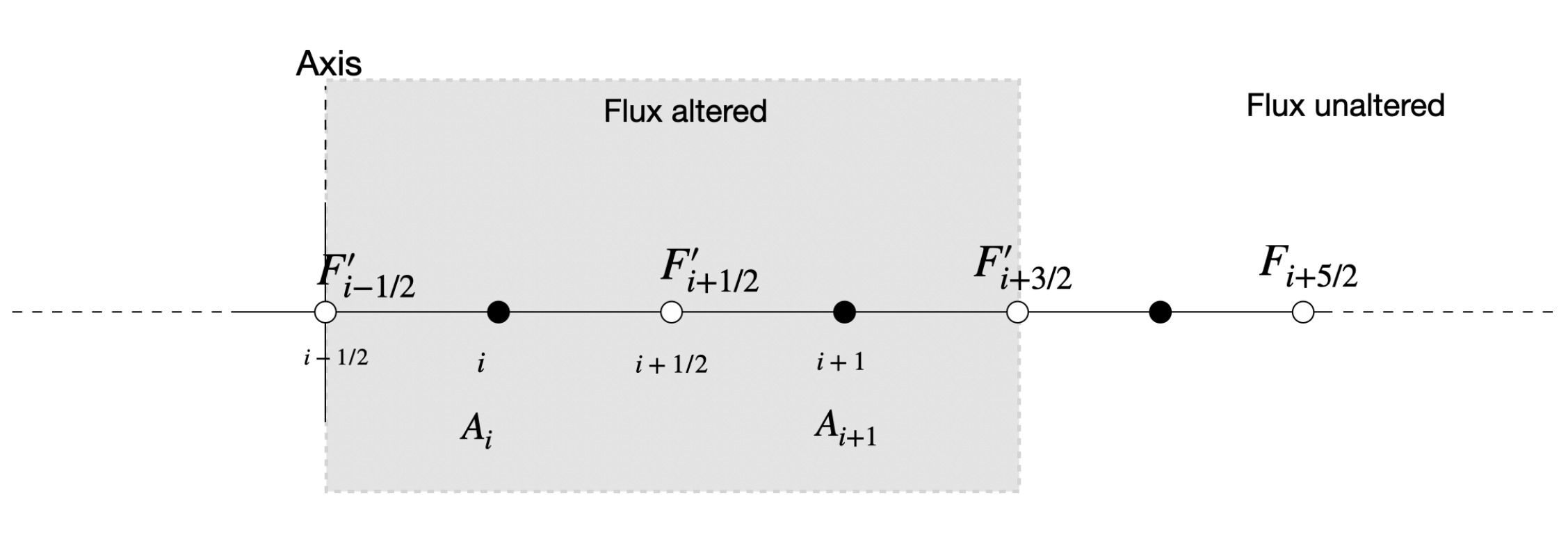}
    \caption{Schematic illustrating the modified conservative implementation.  Quantities labeled by $A$ denote conservative variables; those labeled by $F$ denote the corresponding fluxes. Grid points marked with filled circles represent cell-center, and grid points with open circles represent cell-faces. The fluxes are altered for the first three layers of cell-faces that correspond to two layers of cell-centers}
    \label{SchematicModCons}
\end{figure}
Due to axisymmetry and regularity, the flux on the axis is zero. However in implementation, owing to the choices in the reconstruction, the fluxes on the cell face corresponding to the symmetry axis might not exactly vanish due to reconstruction truncation error. As a result the divergence close to the axis would also comprise a truncation error, that as discussed in Section \ref{Conservative_Scheme} may lead to an accumulation of conserved quantities near the axis. However, it is always possible to choose the reconstruction method in a way that enforces zero flux on the axis. In SpEC, we require $u_i$ on cell faces for calculating fluxes.  In~\cite{Jerred}, variables reconstructed included factors of $\varpi$ to correspond to flat-space orthonormal components, meaning that $u_\phi/\varpi$ was a reconstructed variable (rather than, say, reconstructing $u_\phi$ directly).  Regularity demands $u_\phi/\varpi=0$ on the symmetry axis, but a reconstructed value will be zero only to within truncation error, leading to non-zero axis flux.  A preferable procedure is to
reconstruct $u_\phi/\varpi^2$ and multiply pointwise by $\varpi$ on each face to obtain face values of $u_\phi/\varpi$. The advantage of this can be understood in two ways: 1. Multiplying by $\varpi$, which is exactly zero on the axis, enforces the regularity condition. 2. In the Newtonian limit, $u_\phi/\varpi^2 \sim \Omega$.  Near the axis, regularity demands $\partial\Omega/\partial\varpi\rightarrow 0$, and a nearly constant function can be interpolated with small error.

As a further precaution, the divergence close to the axis could also be altered such that it is well-behaved. Note that the cancellation of flux contributions to the change in $M_A$ which makes flux-conservative schemes conservative is independent of the values of the face fluxes.  The conservation would still be maintained to roundoff error even if the fluxes were to be altered. We use this freedom to adjust fluxes close to the axis such that their divergence is well behaved. 

Analytically, the divergence of a flux is the same whether it is calculated in factored or conservative form, but the finite difference approximations of the two differ for finite resolution.  Starting from the conservative scheme, we alter the flux of each source-free evolution variable on cell faces close to the axis such that the divergence of the flux on the cells enclosed by these faces is the same as for the factored scheme:
\begin{equation} \label{DivCondn}
\begin{split}
    \partial_x \hat{F}^x &= \partial_x(\varpi^n\tilde{F^x}), \\
                    &= \varpi^n\partial_x \tilde{F^x} + n\varpi^{n-1}\frac{\partial \varpi}{\partial x}\tilde{F^x}\ . \\
\end{split}
\end{equation}
Here $\hat{F}^x$ denotes the flux in this new method that we call ``modified conservative'', and $\tilde{F^x}$ denotes the factored flux. This, upon solving for $\hat{F}^x$, gives
\begin{equation} \label{ModConsFlux}
    \begin{split}
        \hat{F}^x_{i+1/2} &= \varpi_i^n ( \tilde{F}^x_{i+1/2} - \tilde{F}^x_{i-1/2} )      \\
                      &+ n \Delta x^i \varpi_i^{n-1} \mathcal{J}^1_{1 i} \text{ }  {\tilde{F}^x_i}           \\
                      &+ \hat{F}^x_{i-1/2},                \\
    \end{split}
\end{equation}
where $\mathcal{J}^1_{1 i} = {\partial \varpi}/{\partial x}$ (stored as a component of the local-to-global map Jacobian) is evaluated at the cell center.

\begin{figure}[h!]
    \centering
    \includegraphics[width=1\linewidth]{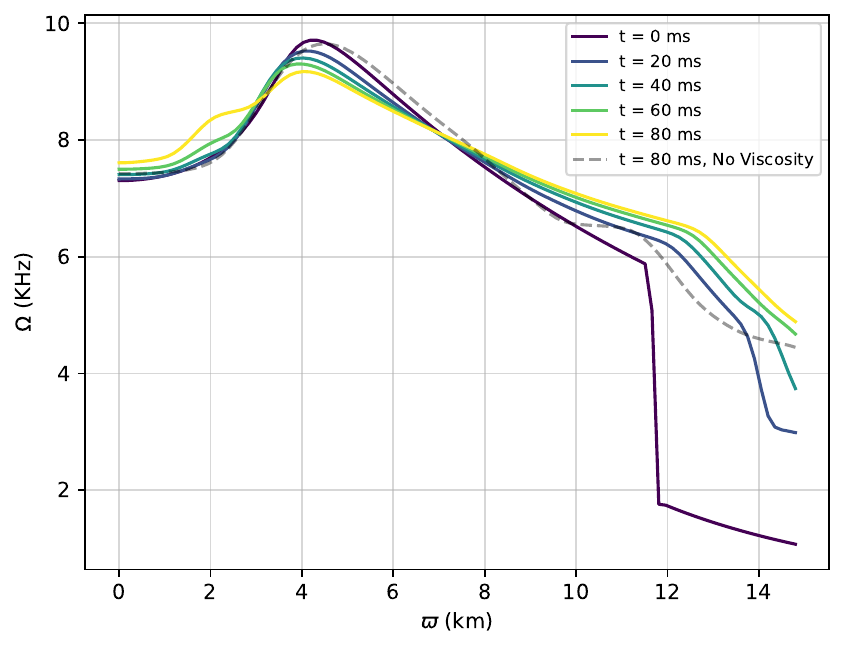}
    \caption{Angular velocity plotted on the equator for the remnant at different times evolved for a longer time using the modified conservative scheme. The black-dashed line shows the velocity profile for a remnant evolved with no viscosity.}
    \label{fig:OmegaProfilesLongRun}
\end{figure}

The modified flux on a cell face depends upon the modified flux at the neighboring cell face, $\hat{F}^x_{i-1/2}$, implying that the above equation is a recurrence relation for the new fluxes.  It therefore requires that one face flux layer be specified.  Fortunately, the first layer, 
$\hat{F}^x_{-1/2}$, lies on the axis, and so regularity together with the assumed symmetry across the axis demands that $\hat{F}^x_{-1/2}=0$, which we enforce.  One can then calculate $\hat{F}^x_{1/2}$ using Eq~(\ref{ModConsFlux}), and from this $\hat{F}^x_{3/2}$, and so forth.  We choose to alter the fluxes for the first three layers of the cell-faces next to the coordinate singularity(see Fig. \ref{SchematicModCons}). As a consequence this would also alter the divergence for the third-layer of the cell-centers. However, such changes that are proportional to the grid resolution $\Delta x$, would be small and insignificant as also supported by smooth angular velocity profiles observed in Fig.(\ref{fig:OmegaProfilesLongRun}).  

We plot the drift $\delta X$ in global quantities for modified conservative scheme in Fig.~\ref{fig:TotalMassAndAngMomError} along with conservative and factored scheme. For all cases, the drift magnitude increases with time, with the drift being negative for angular momentum (i.e. angular momentum is lost). For conservative and modified conservative schemes, the drift is about 3 orders of magnitude smaller than with the factored scheme. The difference in errors could be attributed to maintaining flux-conservative form in conservative schemes as opposed to the presence of an effective source term in the factored scheme that introduces truncation error. We find no dependence of the accuracy of conserved quantities on the grid resolution for modified conservative scheme unlike factored scheme for which the accuracy changes with the grid resolution.

Although the drift in conservative schemes is much smaller than with the factored scheme, the errors remain significantly larger than the expected roundoff error, and they accumulate with time.  This is due to adjustments to evolution variables in the low-density ``atmosphere'' region outside the star and (to a lesser extent) to material entering the atmosphere from the boundary.  These simulations impose a density floor and a velocity ceiling at low threshold densities; we find the drift to be sensitive to these threshold values. The angular velocity profiles for the modified conservative scheme are smooth close to the axis, similar to what is seen for factored scheme, see Fig.~(\ref{fig:OmegaOnLine}). To summarize, the modified conservative scheme conserves global quantities with same accuracy as the flux-conservative scheme, and eliminates non-smooth features close to the axis. 

We further evolve the remnant for 80\,ms to test the modified conservative scheme for long-term stability.  At all times, the angular velocity profile maintains smoothness close to the axis, as shown by the profiles plotted in Fig.~\ref{fig:OmegaProfilesLongRun}. At $t=0$, the rotation profile is non-monotonic, with the peak($\Omega \sim 9$\,kHz) located at $\varpi\approx4$\,km. As the star evolves, the angular velocity profile changes and the peak spreads out, decreasing in magnitude. The angular velocity decreases for $3\,\text{km} \lesssim \varpi \lesssim 7\,\text{km}$, and increases for $\varpi \lesssim 3\,\text{km}$ and $\varpi \gtrsim 7\,\text{km}$, indicating angular momentum transport from the peak to other regions, as expected for the MRI turbulent transport model.

Truncation error could lead to numerical viscosity that could alter the angular velocity profiles and may dominate over the MRI model.  This would indicate that higher resolution is needed to resolve the physical transport, which would be an obstacle to the goal of qualitatively correct simulations at low resolution.  To check for this possibility, we evolve the remnant without viscosity and determine the effect of numerical viscosity on the angular velocity profiles. As shown in the figure, for the case with no viscosity, the remnant roughly maintains the rotation profile with no significant angular momentum transport. The angular velocity increases slightly for the region inside the star that is close to the surface (and thus rather low density).  It is unsurprising that stationarity is not maintained as well in this region, since nonsmoothness at the surface always lowers accuracy, and this layer is convectively unstable.

\subsection{Comparison with other methods}\label{OtherMethods}

The cartoon method~\cite{Alcubierre:2001} is a technique for evolving an axisymmetric system, wherein the hydrodynamic equations are solved in Cartesian coordinates only on one of the many meridional planes.  For a system which is axisymmetric about the $z$-axis, the hydrodynamic equations are evolved on the $x-z$ ($y$=0) plane alone.  The axisymmetry of the system, $\mathcal{L}_{\partial/\partial\phi}T=0$, makes it possible to obtain the solution in any other plane. The value of a field variable at any ($x,y,z$) point could be obtained through a combination of rotation of the $x-z$ system and interpolation of the field quantities. The perpendicular derivatives, in this case the $y$-derivatives which do not necessarily vanish for non-scalar field variables, are calculated from the values of the field variables in adjacent planes.  However, obtaining fields on adjacent planes by interpolation leads to loss of accuracy in conserved quantities. Analytic formulae for the perpendicular derivatives, $\partial_y$, have been derived, in which the derivatives are expressed in terms of field variables in the x-z plane, $\partial_yF^y(x,0,z) = \mathcal{S}(x,z)$~\cite{Pretorius:2005,Hilditch:2015}. The flux-conservative equation in the $x-z$ plane can be rewritten as
\begin{equation}\label{CartoonAfterHilditchFormula}
\begin{split}
     \partial_t A(x,0,z) + \partial_x F^x(x,0,z) & \\
     + \partial_z F^z(x,0,z) + \partial_y F^y(x,0,z) &= 0, \\
     \implies \partial_t A + \partial_x F^x(x,0,z) + \partial_z F^z(x,0,z) &= -\mathcal{S}(x,z), \\
\end{split}
\end{equation}
where the perpendicular derivatives on the RHS appears as an effective source term, one that depends pointwise on the field values in the $x$-$z$ plane. Since the equations are evolved in Cartesian coordinates, the axis issues due to coordinate singularities are avoided.  However, since the equations are in non-conservative form, they conserve global quantities only up to truncation error~\cite{Shibata:2003}.

Numerical evolution of the Einstein-hydrodynamics system in curvilinear coordinates has been carried out in the reference-metric formulation~\cite{BaumgarteEtAl:2013,MonteroEtAl:2014,BaumgarteMoneteroMüller:2015}, and from this one can naturally obtain a 2D formulation by choosing a polar coordinate system and setting azimuthal derivatives to zero.  Here, one introduces a reference 3-metric $\hat{\gamma}_{ij}$, which is the flat-space metric in the chosen coordinate system [$\hat{\gamma}_{ij}=$diag$(1,1,\varpi^2)$ for cylindrical-polar coordinates.]  Source-free conservative equations can be written as
\begin{equation}\label{ReferenceMetricFormalism}
    \partial_t A + \hat{\mathcal{D}}_j f^j = 0,
\end{equation}
where $\hat{\mathcal{D}}$ is the covariant derivative with respect to the reference metric.  This covariant derivative can be calculated in one of two ways.  First, one can expand the covariant derivative in terms of partial derivatives and connection coefficients.  Then the above conservative equation becomes
\begin{equation}\label{ReferenceMetricFormalism2}
    \partial_t A + \partial_j f^j = - \hat{\Gamma}^k_{jk} f^j,
\end{equation}
where $\hat{\Gamma}^k_{jk} f^j$ is a geometric source term that need not be zero. The coordinate singularities are handled by the choice of reference-metric and dealt with analytically, but the evolution equations are now cast in a non-conservative form, and do not conserve global quantities up to roundoff error\cite{BaumgarteEtAl:2013,MonteroEtAl:2014,BaumgarteMoneteroMüller:2015,Mewes:2020,Terrence:2024}.  This method of evaluating fluxes is clearly equivalent to what we have called the ``factored scheme'' above.

The alternative method for calculating covariant derivatives with respect to the reference metric is to utilize identities of the form
\begin{eqnarray}
    \hat{\mathcal{D}}_b f^b &=& \hat{\gamma}^{-1/2}\partial_b\left(\hat{\gamma}^{1/2}f^b\right)\\
    \hat{\mathcal{D}}_b T_a{}^b &=& \hat{\gamma}^{-1/2} \partial_b\left(\hat{\gamma}^{1'2} T_a{}^b\right) - \hat{\Gamma}^c{}_{ba} T_c{}^b .
\end{eqnarray}
Note $\hat{\gamma}^{1/2}=\varpi$. This is how the reference metric covariant derivative is calculated, for example, in~\cite{Lam:2025,Patrick:Gmunu,Patrick:2024}.  For polar coordinates, the connection coefficients vanish for conservative variables $\hat{\Gamma}^c{}_{ba} = 0$ and thus lead to a conservative form of the equation. To summarize, the reference metric formalism does not provide an alternative to the conservative vs. factored choice, but it does provide a geometric understanding of this choice.

Note that the conservative implementation in~\cite{Lam:2025,Patrick:Gmunu,Patrick:2024} differs with the implementation in~\ref{Conservative_Scheme} in how the differential equations are discretized. In~\cite{Lam:2025,Patrick:Gmunu,Patrick:2024}, the hydrodynamic equations are solved using finite-volume methods as against finite-difference method used in section~\ref{Conservative_Scheme}. The spatially discretized form of a 2D flux-conservative equation on a uniform grid with grid spacing $\Delta x, \Delta y$ at gridpoint \textit{i,j} for a finite-volume conservative scheme is,
\small
\begin{equation}\label{FVConservativeFluxForm}
    \begin{split}
         \partial_t \langle A \rangle _{i,j} =& \langle S \rangle_{i,j}\ + \Delta V^{-1} \times \\
                            & \bigl\{ [ \langle F \rangle_{i-1/2,j}^x \Delta A_{i-1/2,j}^x - \langle F \rangle_{i+1/2,j}^x \Delta A_{i+1/2,j}^x ] \\
                            & + [ \langle F \rangle_{i,j-1/2}^y \Delta A_{i,j-1/2}^y - \langle F \rangle_{i,j+1/2}^y \Delta A_{i,j+1/2}^y ] \bigr\} \\
              \end{split}
\end{equation}
\normalsize
where $\Delta V$ and $ \Delta A_{i,j}^k$ are the volume and the surface area of the cell $i$, $\langle A \rangle _{i,j}$, $\langle S \rangle_{i,j}$ are the volume average of the corresponding quantities, and $\langle F \rangle_{i,j}$ are the surface-averaged quantities of the flux terms at the cell interfaces. While both finite-difference and finite-volume conservative schemes conserve global quantities up to roundoff error, finite-volume schemes have an added advantage of the fluxes through the axis vanishing automatically ($\Delta A_{i,j} = 0$ for the cell-face on the axis). Thus the divergence near the axis is dealt with by construction for finite-volume conservative schemes, and these seem to exhibit smoothness in angular velocity profiles close to the axis~\cite{Patrick:2024}.  Nevertheless, finite difference methods remain a commonly used discretization method in computational astrophysics; they allow relatively straightforward implementations in multiple dimensions and at high order, so techniques for ensuring good behavior near the axis remain well worth pursuing.

\section{Entropy Density Evolution}{\label{EntropyEvolution}}

The gas in post-merger remnants undergoes heating and cooling via viscosity and neutrino-matter interactions, but at the 
resolutions commonly used in the post-merger simulations, artificial heating by numerical truncation error might be comparable to these physical effects. If hydrodynamic fields can be assumed to be smooth, it is acceptable to directly evolve the entropy density of the fluid.  Such evolution will, in fact, be much more accurate in supersonic and degenerate regions, where thermal energy is a subdominant contribution to $\tau$, and numerical error can easily overwhelm the thermal contribution.  Fortunately, shocks are not expected to develop during the secular post-merger evolution of our isolated hypermassive star.  In more realistic post-merger states, shocks might arise due to fallback accretion or steepening of acoustic modes of an initially perturbed state.  In such cases, it would be necessary to identify points where shock or reconnection heating might occur, either using local gradients to identify discontinuities or by some a priori expectation for where they might occur.  Below, we show how the source term for the entropy density can be easily calculated from the source terms for other conservative variables.

Consider a fluid with number density $n$, energy density $\rho$, pressure $P$, specific entropy $s$, and 4-velocity $u^{\alpha}$.  The first law of thermodynamics for a fluid element can be written
\begin{equation}
  \frac{d\rho}{d\tau} = \frac{\rho+P}{n}\frac{dn}{d\tau}
  + nT\frac{ds}{d\tau} + n\sum_i\mu_i\frac{dY_i}{d\tau}\, 
\end{equation}
where $d/d\tau\equiv u^{\alpha}\nabla_{\alpha}$ is the proper time derivative, and $Y_i=N_i/N$ is
the number fraction (not mass fraction) of particle species $i$.

The perfect fluid stress tensor is
\begin{equation}
  T_F^{\mu\nu}=(\rho+P)u^{\mu}u^{\nu}+Pg^{\mu\nu}
\end{equation}
A quick calculation shows
\begin{equation}
  u_{\beta}\nabla_{\alpha}T_F^{\alpha\beta}=-\frac{d\rho}{d\tau}+\frac{\rho+P}{n}\frac{dn}{d\tau}
\end{equation}
Therefore
\begin{equation}
  n\frac{ds}{d\tau} = \nabla_{\alpha}(nsu^{\alpha})=\frac{1}{T}
  \left[-u_{\beta}\nabla_{\alpha}T_F^{\alpha\beta}
    -n\sum_i\mu_i\frac{dY_i}{d\tau}\right]
\end{equation}
Our entropy evolution variable is $\rho_{\star}s\equiv nW\sqrt{\gamma}s$, where
$W=\alpha u^0$ is the Lorentz factor.  Then  $\rho_{\star}s$ obeys a continuity equation with source term
\begin{equation}
  (\rho_{\star}s)_{\rm RHS} = \alpha\sqrt{\gamma}\nabla_{\alpha}(nsu^{\alpha})
\end{equation}
Expand the 4-velocity as follows:
\begin{equation}
  u_{\beta}=Wn_{\beta} + Wv_{\beta}
\end{equation}
where $n_{\mu}=(-\alpha,\mathbf{0})$ is the normal vector and
$v^{\mu}n_{\mu}=0$, implying $v^0=0$.  Note that $v^{\mu}$ differs from
the transport velocity $V^{\mu}=u^{\mu}/u^0$. Then
\begin{equation}
    \begin{split}
      u_{\beta}\nabla_{\alpha}T_F^{\alpha\beta}
      =&  W n_{\beta} \nabla_{\alpha}T_F^{\alpha\beta}
      + Wv^{i}\nabla_{\alpha}T_F^{\alpha}{}_{i}     \\
      =& W\nabla_{\alpha}(n_{\beta}T_F^{\alpha\beta})
      - W\nabla_{\alpha}n_{\beta}T_F^{\alpha\beta} \\ 
      \text{ }& + Wv^{i}\nabla_{\alpha}T_F^{\alpha}{}_{i}
  \end{split}
\end{equation}
These terms can be identified with the source terms of the conservative energy and momentum evolution variables:
\begin{eqnarray}
  \tau_{\rm RHS}
  &=& -\alpha\sqrt{\gamma} n_{\mu}\nabla_{\nu}T_F{}^{\mu\nu} \\
  &=& \partial_t\tau + \partial_i F_\tau^i 
  + \alpha\sqrt{\gamma}T_F^{\mu\nu}\nabla_{\nu}n_{\mu} \\
  (S_i)_{\rm RHS} &=& \alpha\sqrt{\gamma} \nabla_{\mu} T_F^{\mu}{}_i
\end{eqnarray}
Finally, the chemical potential term.  We have three particle species:  electron/positron leptons (``e''), protons (``p'') and neutrons (``n'').  Under beta
reactions, $dY_e=dY_p=-dY_n$.  Then
\begin{equation}
  \sum_i\mu_i\frac{dY_i}{d\tau}
  = (\mu_e + \mu_p - \mu_n)\frac{dY_e}{d\tau}
\end{equation}
The derivative $dY_e/d\tau$ is related to our composition evolution variable $\rho_{\star}Y_e$:
\begin{equation}
n\frac{dY_e}{d\tau} = \nabla_{\alpha}(nY_eu^{\alpha}) = \alpha\sqrt{\gamma}(\rho_{\star}Y_e)_{\rm RHS}
\end{equation}
Putting this altogether,
\begin{equation}
\begin{split}
        (\rho_{\star} s)_{\text{RHS}} =& \frac{1}{T} \Big[ 
     W (\tau)_{\text{RHS}} -W v^i (S_i)_{\text{RHS}}  \\
    & - (\mu_e + \mu_p -\mu_n)(\rho_\star Y_e)_{\text{RHS}} \Big] \\ 
\end{split}
\end{equation}

Like several other relativistic hydrodynamics codes, SpEC already evolves an auxiliary entropy variable $\rho_\star s$ via a continuity equation (ignoring heating) intended to be used, along with $\rho_\star$ and $S_i$, to calculate primitive variables $W$ and $T$ at points where recovery using ($\rho_\star$,$\tau$,$S_i$) is deemed to have failed--the root-finding process either finding no root or finding a physically unacceptable one.  At the end of the step, both $\tau$ and $\rho_\star s$ are overwritten using the chosen $T$ and $W$, which will fix an unacceptable $\tau$ or communicate heating information to $\rho_\star s$.  Implementation of entropy evolution is done simply by adding the heating source term to the $\rho_\star s$ evolution equation and changing the logic of primitive variable recovery to always use the entropy variable.  We have tested our implementation by creating arbitrary artificial ($(\tau)_{\text{RHS}}$, $(\rho_\star Y_e)_{\text{RHS}}$, $(S_i)_{\text{RHS}}$), evolving a single step once using $\tau$ and once using $\rho_\star s$, and comparing increments of the fluid variables.

\begin{figure}[h!]\label{EntropyComparison}
    \centering
    \includegraphics[width=1\linewidth]{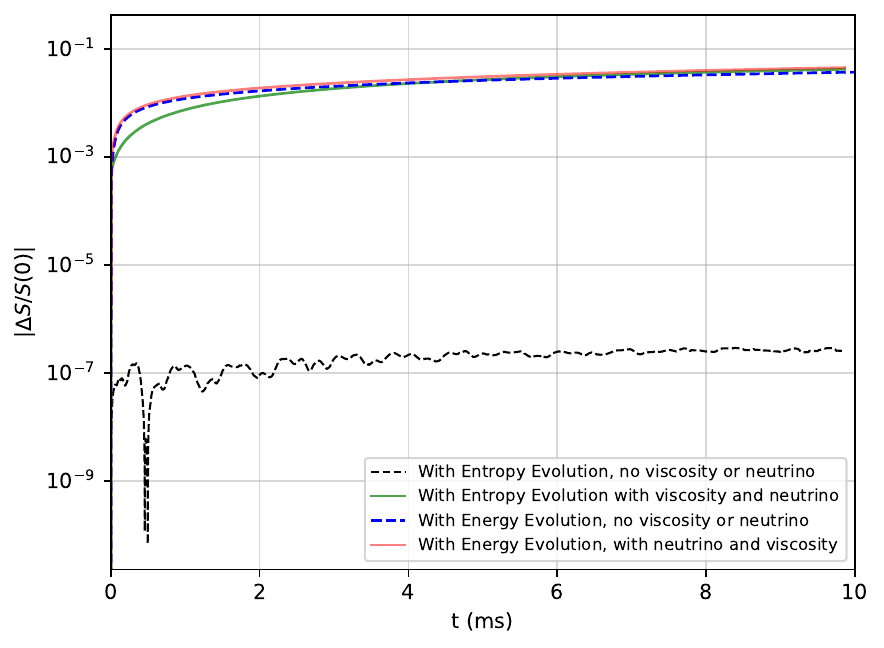}
    \caption{Relative change of total fluid entropy from its initial value for different schemes as a function of time for different methods and included physics. The relative difference is negative for the case with entropy evolution and no viscosity or neutrino effects (i.e. the star slightly cools).}
    \label{fig:TotalEnt}
\end{figure}

In Fig.~\ref{fig:TotalEnt}, we plot the difference of the total entropy of the fluid from $t=0$, $\Delta S_F(t) = S_F(t) - S_F(0)$, for the system evolved using the entropy density and compare it with the cases where instead the energy density $\tau$ is evolved. We further include cases with no viscosity (i.e. perfect fluid) or matter-neutrino interaction, for which the evolution should be adiabatic.We find all the evolutions with entropy density to be stable along with an improvement in performance with much smaller CPU time per code time-step. For the cases where the system is evolved with energy density, artificial heating dominates over heating and cooling via viscosity and neutrino-matter interaction, as evident by similar $\Delta S_F$ values with and without viscosity and matter-neutrino interaction. In the absence of matter-neutrino interaction and viscosity, $\Delta S_F$ for the system evolved using entropy density is approximately four orders of magnitude smaller than the system evolved using energy density, indicating that entropy density evolution effectively suppresses artificial heating.

We remark that $\Delta S_F$ with entropy evolution, although small, is still larger than roundoff error.  Atmosphere corrections to density and temperature can play a small role in this, but the main cause of this error is a peculiarity in our implementation of the entropy evolution.  After $\rho_\star s$ is evolved forward a time substep, it is used to infer $T$ via root finding.  At the end of each timestep, evolution variables are recomputed using primitive variables like $T$ (to communicate atmosphere fixes to the evolution variables).  Thus, a finite tolerance in the root finder for $W$ and $T$ creates error in $\rho_\star s$.  For example, a slight adiabatic compression might produce an increase in $T$ below the root-finder tolerance, so the original $T$ will be retained and the point effectively evolving isothermally. We find that altering the root solver parameters like tolerance and maximum number of allowed iterations leads to significantly better entropy conservation with an order of magnitude lower $\Delta S_F$ values.  The computational cost is insignificant, but we report our original entropy evolution simulations because we consider the entropy conservation already acceptable (i.e. artificial changes insignificant compared to the physical effects described below). 

For evolutions that include the effect of viscosity and matter-neutrino interaction, the cases with and without entropy density evolution have comparable $\Delta S_F$ values, with the former somewhat lower for the first few milliseconds.  This indicates that, for evolutions with energy variable $\tau$ at our chosen resolution, artificial heating dominates over physical heating for the first milliseconds, but that accummulated physical heating dominates thereafter, a reassuring result for modest-resolution energy-variable evolutions.  Convergence tests also confirm that total entropy for simulations with $\tau$ converge toward total entropy for simulations with $\rho_\star s$, the latter being almost resolution-independent.

Although not shown here, we have also performed an evolution of the system using entropy density that includes matter-neutrino interaction but without viscosity. For this case, we observe negative $\Delta S_F$ values, i.e. the fluid entropy decreases with time. Note that this is not a violation of the second law of thermodynamics since the total entropy of the system would be a sum of the fluid entropy and neutrino entropy.  Thus, the fluid entropy loss (cooling) due to energy transfer from the fluid to the neutrinos is larger than fluid entropy gain due to thermalization of fluid in different regions of the star by neutrino energy transport. 

\section{Summary and Conclusion}\label{Conclusion}

The value of 2D axisymmetric simulations of systems such as post-merger binary neutron star remnants is that they can probe late times or many cases quickly and cheaply.  Their realism is intrinsically limited by the neglected azimuthal information:  by limitations of the turbulent transport model and by the neglect of nonaxisymmetric modes.  There is, then, little reason to pursue very high accuracy in 2D simulations at the cost of speed.  However, errors that accumulate, i.e. drifts in extensive quantities, could easily cause qualitative differences from the true behavior at late times.  The promise of conservative schemes is to completely remove such drifts even in otherwise inaccurate runs.  Necessarily, this means global conservation laws cease to be useful as code checks, but the gain in late-time qualitative realism is generally considered worth it.  However, 2D conservative schemes can also come with a trade-off in robust smoothness near the axis, and schemes that deal with the axis coordinate singularity sacrifice smoothness.  We have presented methods to provide both near-axis smoothness and exact (up to atmosphere and boundary effects) conservation of baryon mass and angular momentum.  We also implement an entropy evolution scheme for problems with exclusively smooth flows.

Axisymmetric SpEC evolutions can now evolve for second timescales at modest resolution without artificial drift errors.  We will next use this to probe post-merger systems for the timescales needed (often multiple seconds) to study postmerger-driven electromagnetic counterparts to the merger's gravitational wave signal.  We have implemented in this paper a method to generate equilibria with properties (mass, spin, entropy, and features of the rotation profile) matching that of post-merger putcomes from 3D simulations. We can easily do this to extend many 3D simulations or to create hypothetical remnants differing in controlled ways, in order to probe the sensitivity of the subsequent evolution to the initial state.

Another available avenue for exploration is the sensitivity to the model of turbulent transport.  As we explain in the Appendix, much uncertainty remains regarding MHD instabilities and their saturation properties in the deep interior of differentially rotating neutron stars.  A great deal of guesswork was needed even in attempting to create a qualitatively correct model.  As we explain, there are reasons to question whether commonly used models are reliable even qualitatively in general cases.  Our proposed model addresses some of these worries but remains uncalibrated.  High resolution 3D MHD simulations are crucial for resolving uncertainties in the feedback of nonaxisymmetric turbulence onto the azimuthal mean flows.  A complementary study could use 2D simulations to vary the assumed turbulence model to determine the sensitivity of the evolution to its unknown parameters.

\begin{acknowledgments}

We thank Conor Bartol for analyzing initial data.
M.D. gratefully acknowledges support from the
NSF through grant PHY-2110287. M.D. and F.F. grate-
fully acknowledge support from NASA through grant
80NSSC22K0719.  F.F. gratefully acknowledges support
from the Department of Energy, Office of Science, Of-
fice of Nuclear Physics, under contract number DE-
SC0025023 and from the NSF through grant AST-
2107932.  P.C.-K.C. acknowledges support from
NSF Grant PHY-2020275 (Network for Neutrinos,
Nuclear Astrophysics, and Symmetries (N3AS)).

\end{acknowledgments}

\appendix
\section{Turbulence Modeling by Viscosity}\label{TurbulenceModeling}

The post-merger remnant is subject to several magnetohydrodynamic (MHD) instabilities, in particular the magnetorotational instability (MRI) and the Tayler-Spruit instability (TSI), which can cause turbulence inside the star. The turbulence can transport angular momentum inside the star, thereby affecting the state of the remnant on a secular timescale. The angular momentum transport could also be reproduced using subgrid models that incorporate unresolved small scale physics in terms of large-scale quantities~\cite{Miravet-Tenes:MRIModel,Miravet-Tenes:KHModel,Miravet-Tenes:2025}. These transport effects could also be modeled using an effective viscosity, where the viscosity is specified by a characteristic mixing length $\ell\equiv \nu/c_s$, where $\nu$ is the effective kinematic viscosity, and $c_s$ is the fluid adiabatic sound speed. Many simulations use an alpha viscosity model in disks~\cite{ShakuraSunyaev:1973}, according to which the mixing length is given by
\begin{equation}
    \ell =\frac{\alpha c_s }{\Omega},
\end{equation}
where $\alpha$ is a dimensionless parameter that is expected to be less than one.  MHD instabilities can exist inside stars as well, but the resulting angular momentum transport will often not be modeled accurately using the same alpha disk prescription.  A realistic model should take into account the criteria that determine the regions in which a given MHD instability and accompanying transport operate.  Below, we discuss instability criteria for MRI and TSI instabilities and different effective viscosity models for their transport effects on the mean stellar flow.  For this discussion, we define the cylindrical radius $\varpi$, the effective kinematic viscosity $\nu$, and the Alfven timescale $\omega_A\equiv v_A/\varpi\sim B/(\varpi\rho^{1/2})$.

\subsection{MRI-driven turbulence}

Ignoring buoyancy effects, the MRI instability criterion for a weakly magnetized conducting fluid is~\cite{BalbusHawley:1998}
\begin{equation}\label{SpecificMRICriterion}
    q < 0,
\end{equation}
where,
\begin{equation}\label{RadialShear}
q \equiv \frac{d \ln\Omega}{d\ln r},   
\end{equation}
is the local radial shear. Thus, the system is MRI unstable whenever the angular velocity decreases outward. 

Many simulations show that post-merger remnants have non-monotonic rotation profiles~\cite{Kastaun:2014fna,Hanauske:2016gia,de2020numerical,iosif2021equilibrium,iosif2022models,cassing2024realistic,staykov2023differentially}; the angular velocity increases with the radius in the core and decreases in the envelope.  The core, then, is MRI-stable and is expected to have weak angular momentum transport (at least, if local MRI instability is the only source of turbulence). This is consistent with high-resolution GRMHD simulations that report small effective $\alpha$ values in inner regions~\cite{Kiuchi:2018}. Thus, constant $\alpha$-disk model are unrealistic when modeling post-merger NS remnant.

In \cite{RadiceMixLength:2020, RadiceMixLength:2023}, Radice and Bernuzzi devised a numerical relativity calibrated transport model, that we refer here as \texttt{DF1}~\cite{RadiceMixLength:2020} and \texttt{DF2}~\cite{RadiceMixLength:2023}. In these models the mixing length $\ell$ is taken to be a fixed function of density, $\ell = \ell(\rho)$ where the density dependence is obtained by numerical fit to the density-binned $\alpha$ values from Kiuchi~{\it et al.}'s GRMHD merger simulations~\cite{Kiuchi:2018,Kiuchi:2023}. For \texttt{DF1}, the mixing length has a fixed non-monotonic dependence upon density and diminishes for low and high densities, whereas for \texttt{DF2} the mixing length is monotonic and diminishes for high densities. Both these models account for the MRI-stable region at high densities and are a significant improvement upon a constant $\alpha$ model.  However, the density at which the star switches between MRI-stable and unstable can change during the seconds of post-merger evolution. The transition density will also vary across systems; binaries with different NS masses, mass ratios and pre-merger spins will produce somewhat different remnants. Also, the effective viscosity changes in strength as the star evolves, as seen by the time-varying $\alpha$ values reported in \cite{Kiuchi:2018}.  Thus, a mixing length with fixed dependence upon density might not accurately capture angular momentum transport across all remnants and for all phases across the remnant's evolution. 

Some 1D stellar simulations incorporating MRI transport utilize an effective viscosity with a dependence on the local radial shear, $\nu = \nu(q)$.  
The functional form of such models is motivated by arguments in~\cite{Spruit:2001tz,Wheeler:2015} outlined below.

The effective kinematic viscosity, $\nu_{\text{mag}}$,  is set so that the resulting stress $T_{r\phi}$ matches the average Maxwell stress $S$ from the turbulent magnetic field:
\begin{equation}\label{AzimuthalMaxwellStress}
    S \equiv \frac{\langle B_r B_\phi\rangle}{4 \pi} = \rho q \Omega \nu_{\text{mag}}\ ,
\end{equation}
which, upon solving for $\nu_\text{mag}$, gives $\nu_{\text{mag}}=S/\rho q \Omega$. Wheeler \textit{et al.}~\cite{Wheeler:2015} define a stress efficiency parameter $\alpha_{\text{mag}}$ as
\begin{equation}\label{AlphaWKC}
    \alpha_{\text{mag}} \equiv \frac{\langle B_r B_\phi\rangle}{P_0}\ ,
\end{equation}
where $P_0$ is the magnetic pressure of the saturated field. Writing $S$ in terms of the stress efficiency parameter gives
\begin{equation}\label{NuMag2}
        \nu_{\text{mag}} = \alpha_{\text{mag}} \frac{P_0}{\rho q \Omega}.
\end{equation}
Deep inside the star, the saturation field for the MRI can be estimated to order of magnitude assuming an equipartition between magnetic and differential rotational kinetic energy:  $\omega_A \sim q \Omega$, which gives, $P_0 \sim \rho (q \varpi \Omega)^2$.  Substituting Eq.~(\ref{NuMag2}) for $P_0$ gives
\begin{equation}\label{NuMag3}
    \nu_{\text{mag}} = \alpha_{\text{mag}} q \Omega \varpi^2\ .
\end{equation}

Margalit~\textit{et al.}~\cite{Margalit:2022} propose a functionally similar effective viscosity model 
\begin{equation} \label{eq:nu_vs_ht}
    \nu_{\text{MRI}} = \alpha_g h_t^2 \Omega q\ \text{,  } q<0
\end{equation}
where $h_t$ is the characteristic coherence length scale of the turbulence and the stress efficiency parameter, $\alpha_g$ is defined as $\alpha_g \equiv <T_{r \phi}>/P_g$ where $P_g$ is the gas pressure, and $T_{r \phi}$ is a component of the total mean stress tensor. Aguilera-Miret \textit{et al.}~\cite{Aguilera-Miret:2024} find the characteristic length scale to be roughly the same order of magnitude as the radius of the star, whereas other simulations suggest that $h_t \sim \varpi/3$ ~\cite{Metzger:2009,Margalit:2022}. We follow the later in estimating the characteristic length scale.  This gives the effective viscosity
\begin{equation}\label{MargalitModel}
    \nu_{\text{MRI}} = \frac{1}{9}\alpha_g \varpi^2 \Omega q.
\end{equation}
Eq.~(\ref{eq:nu_vs_ht}) is valid also in supersonic, disk-like regions, but there equipartition is expected to be between magnetic and internal energy:  $\omega_A\sim c_s/\varpi$, and $h_t\sim\lambda_{\rm MRI}\sim c_s/\Omega$ is the disk height, and $\nu \sim \alpha_g q c_s^2/\Omega$, the standard $\alpha$-disk model.  A general model of $\nu_{\rm MRI}$ accounting for both highly subsonic (deep stellar interior) and highly supersonic (thin disk) regions would require a form of $h_t$ which reproduces both limits.  Note that in both limits, the magnetic energy density approaches the smaller of kinetic and internal energy density, so that $h_t$ is closer to the smaller of the two estimates, suggesting, for example, an interpolating function of the form
\begin{equation} \label{eq:matching_h}
h_t \approx \frac{\varpi c_s}{\varpi\Omega + 3c_s}
\end{equation}
Eq.~(\ref{eq:nu_vs_ht}) with Eq.~(\ref{eq:matching_h}) would probably be the best option for a simulation where all limits are anticipated.  For the remnant systems studied in this paper, the envelope and disk region is geometrically thick, and $h_t\sim\varpi/3$ is a good approximation in the disk as well as a reasonable one in the stellar interior, so we use Eq.~(\ref{MargalitModel}) everywhere.

Although Eq.~(\ref{NuMag3}) and Eq.~(\ref{MargalitModel}) are functionally similar, there are important differences in the way the stress efficiency parameter is defined.  As pointed out in Griffiths~\textit{et al.}~\cite{Griffiths:2022}, the stress efficiency parameter deep inside the star is very different depending on the choice of the normalization pressure, $P_g$ or $P_0$. The $\alpha$ parameter also depends on whether the $T_{r\phi}$ considered is just the Maxwell stress or the sum of Maxwell and Reynold stresses.  However, this latter difference will be small, at least in the disk region where the Maxwell stress dominates over the Reynold stress~\cite{Stone:1996}.

The factor $\alpha$ could also depend on $q$, as indicated by local shearing box simulations that find $\alpha$ increases with local shear~\cite{Abramowicz:1996,Masada:2012}. Simulations with zero net magnetic flux, such as Abramowicz~\textit{et al.}\cite{Abramowicz:1996}, estimate the viscosity parameter $\alpha$ caused by magnetic instability and find it to be approximately proportional to the shear-vorticity ratio, $g_q$, which is approximately $q/2$ for small $q$ values. Other local shearing box simulations, with non-zero net magnetic flux such as Masada et al. \cite{Masada:2012} find $\alpha \propto g_q^{0.44}$. 

\begin{figure}[h!]
    \centering
    \includegraphics[width=1\linewidth]{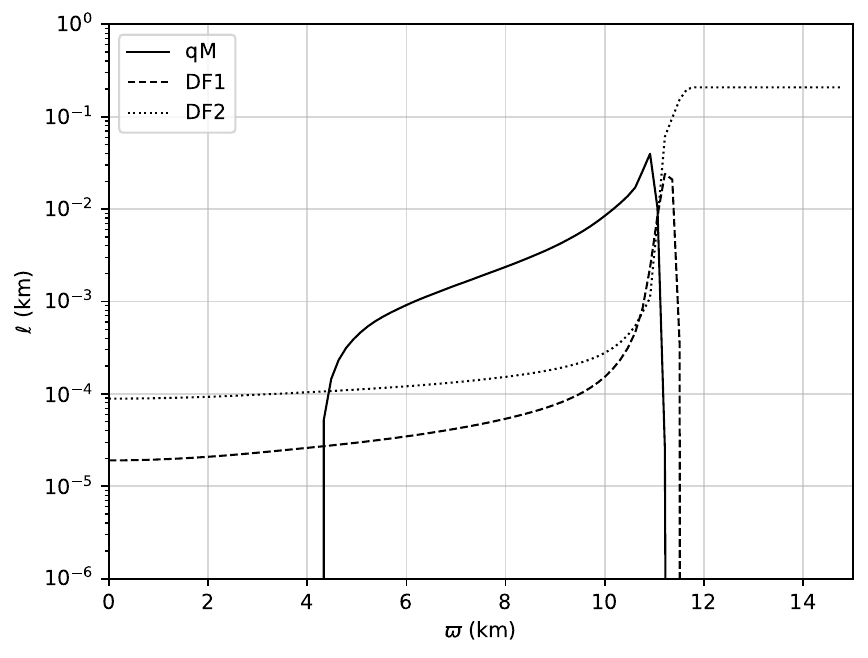}
    \caption{Mixing length profiles for the density-fitted models \texttt{DF1} (dashed line)  and \texttt{DF2}(dotted line) and the q-dependent model \texttt{qM} used in this paper (solid line) with $\alpha = 0.03$, all plotted on the equator of the remnant.}
    \label{fig:MixingLengthComparison}
\end{figure}

We use the effective viscosity given by Eq.~(\ref{MargalitModel}) for $q<0$, with $\nu_{\text{MRI}}=0$ where $q>0$ and exclude the extra $q$-dependence for $\alpha$. We restrict the mixing length to be confined in the star and suppress the mixing length for regions close to the surface, $\tilde \ell (\rho) = \ell(\rho) e^{-0.01(\frac{\rho_{\text{thr}}}{\rho} - 100)}$ for $\rho_\text{thr}/{\rho} > 100$, where $\rho_\text{thr}$ is some threshold density.  In Fig.~\ref{fig:MixingLengthComparison} we plot mixing length on the equatorial line for our q-dependent MRI model, labeled as \texttt{qM} and compare it with Radice and Bernuzzi's density-fitted models \texttt{DF1} and \texttt{DF2}. In the core region, which is MRI-stable (see Fig. \ref{fig:OmegaProfilesLongRun}), the mixing length for \texttt{qM} is zero. For MRI-unstable regions, $\varpi \gtrsim 4\,\text{km}$, the mixing length for \texttt{qM} varies between $0.1 - 100$m. In comparison, the mixing length for \texttt{DF1} and \texttt{DF2} are $\sim 1-2$ magnitudes smaller in much of the MRI-unstable region, and non-zero($\sim 10^{-2}-10^{-1}$m) in the MRI-stable region. The mixing length is confined in the star for \texttt{qM} and \texttt{DF1}, but is non-zero for \texttt{DF2} for regions outside the surface. 

Care should be taken in interpreting this comparison.  Clearly, a major reason for the difference is that our equilibrium star differs from the simulated remnants used to calibrate \texttt{DF1} and \texttt{DF2}.  Perhaps our star, despite being informed by merger simulations as described in Section~\ref{sec:initial_data}, remains unrealistic in important ways.  Even if this is so, the prospect that realistic remnants will deviate significantly from their initial post-merger states motivates future refinements to \texttt{DF1} and \texttt{DF2} to incorporate the effects captured in our model.

Due to the dependence of the strength of the MRI model on $q$, $\ell \propto d \ln \Omega / d \ln r$, any raggedness in the numerical derivative will manifest in the mixing length.  At the initial time, $\Omega$ is smooth, so no such anomalous features are seen in Fig.~\ref{fig:MixingLengthComparison}.  One might worry that numerical error could be amplified during evolution, but we see no indication of this in 80\,ms of evolution.

\subsection{TSI Turbulence Modeling}

In some cases, the TSI can lead to significant momentum transport, perhaps even dominating over the MRI.  The TSI dynamo has been proposed to exist for convectively stable layers where $\omega_A \ll \Omega \ll N$, where $N$ is the Brunt-V{\"a}is{\"a}l{\"a} frequency. Such situations might arise in outer layers of a protoneutron star~\cite{Barrere:2022gwv} or inner (MRI stable) layers of a binary neutron star remnant~\cite{Margalit:2022}.  Simulations indicate appropriate conditions for TSI in binary neutron star mergers, although they cannot yet directly resolve it~\cite{Reboul-Salze:2024jst}.  Simulations resembling a PNS with fallback~\cite{Barrere:2024jyw} support TSI transport with the scaling proposed by Fuller~{\it et al.}~\cite{Fuller:2019ckz}: $\nu_{\rm TSI} \propto r^2 q^2\Omega^3/N^2$. Unlike the MRI, it is nonzero for either sign of $q$.  Margalit~{\it et al.}~\cite{Margalit:2022} find extremely fast (dynamical timescale) damping of differential rotation in neutron stars with this scaling, although using a scaling factor much higher than~\cite{Barrere:2024jyw} suggest.  Most likely, TSI-driven transport is weaker than this, but still important, especially in MRI stable regions as also pointed out in Margalit~{\it et al.}~\cite{Margalit:2022}.

The radial extent of unstable modes is confined to $\ell_r < v_A/N$~\cite{Spruit:2001tz}.  A distinctive feature of TSI is that transport is extremely anisotropic, being much faster in the angular direction (i.e. along pressure contours), for which $\ell_{\perp}\sim r$~\cite{Spruit:2001tz}.  

A model of transport by TSI-turbulence must calculate $N$, deal with the anisotropy and with the $\nu_{\rm TSI}$ scaling in the low-$N$ limit.  Below, we show how these issues are dealt with in our implementation.

For a fluid stratified in direction $\partial/\partial z\equiv -\mathbf{\nabla}P/\left|\nabla P\right|$, Newtonian buoyancy of adiabatically bobbing fluid elements gives (assuming gravity is supported by pressure rather than rotation, i.e. $dP/dz=g\rho$)
\begin{equation}
  N^2 = \left[\left.\frac{d\rho}{dP}\right|_{s,Y_e}  - \frac{d\rho}{dP}\right]\left(\frac{1}{\rho}\frac{dP}{dz}\right)^2
\end{equation}
The adiabatic derivative can be related to the sound speed, and rewriting derivatives in terms of spatial gradients, we have
\begin{equation}
  N^2 = \frac{|\nabla P|^2}{h\rho_0^2c_s^2}
  - \frac{\nabla P\cdot\nabla\rho_0}{\rho_0^2}
\end{equation}

In order to have different $\nu_{\rm TSI}$ parallel to $\mathbf{\nabla}P$ (i.e. across isobars) and perpendicular to $\mathbf{\nabla}P$ (along isobars), introduce the spatial projection operator $P^j_i=r^j r_i/(r^k r_k)$, where $r^i$ can be taken to be $\nabla^iP$ or (since the stellar region is roughly spherical) the coordinate $x^i$.  Then, we generalize $T^{\rm visc}{}_{ij}=-2\rho_0\eta\sigma_{ij}$ to
\begin{equation}
T^{\rm visc}_{ij}
= -2\rho_0\left[\nu_{\perp}\sigma_{ij} + (\nu_{\parallel}-\nu_{\perp})P^k_iP^m_j\sigma_{km}\right]\ .
\end{equation}

Spruit~\cite{Spruit:2001tz} proposes $\nu_{{\perp}\ \rm TSI}\propto N^{-4}$, while Fuller~{\it et al.}~\cite{Fuller:2019ckz} propose $\nu_{{\perp}\ \rm TSI}\propto N^{-2}$.  Low $N$ regions are outside the domain of applicability of their analyses, which assume $N^2>\Omega^2$, but a simulation must nevertheless deal with them somehow.  We regularize $\eta_{\rm TSI}$, the Fuller~{\it et al.}~\cite{Fuller:2019ckz}, as follows:
\begin{equation}
\nu_{{\perp}\ \rm TSI} = \frac{\alpha_{\rm TSI} q^2 r^2\Omega^3}{N^2+\Omega^2}
\end{equation}
In fact, the approximation of a buoyantly confined Tayler instability breaks down in this limit, and only 3D MHD simulations can determine how best to model its transport effects in 2D.  However, as pointed out by Margalit~{\it et al.}~\cite{Margalit:2022}, weak stratification is often a narrow transition region between strong stratification and convective instability, so one can hope that any reasonable treatment of this regime will allow accurate global evolutions.

Only 3D MHD simulations can determine a suitable $\alpha_{\rm TSI}$.  Following~\cite{Barrere:2024jyw}, we use $\alpha_{\rm TSI}=10^{-6}$.  (See Table 1 on page 8, which gives $\alpha=0.01\pm 0.004$ for their scaling parameter, and note that $\alpha_{\rm TSI}=\alpha^3$.)  For this $\alpha_{\rm TSI}$, the TSI mixing length is an order of magnitude smaller than the MRI mixing length (except in the MRI stable region where the latter is zero).  However, even apart from uncertainty in $\alpha_{\rm TSI}$, the calculated $\nu_{\rm TSI}$ for our star is arbitrary because it is in the region where scaling is dominated by the regulator term.

\begin{figure}[h]
    \centering
    \includegraphics[width=1\linewidth]{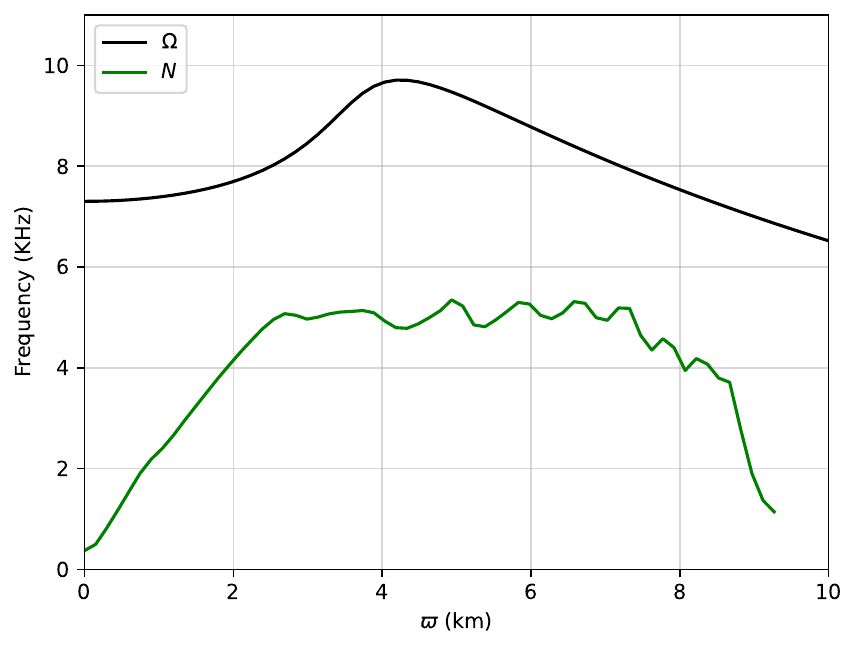}
    \caption{The Brunt-V{\"a}is{\"a}l{\"a} frequency (in green), and angular velocity (in black) of the remnant plotted on the equator at $t=0$. At $\varpi = $ 9.4,\,km the Brunt-V{\"a}is{\"a}l{\"a} frequency becomes imaginary.}
    \label{fig:OmegaVsBVFreq}
\end{figure}

In Fig.\ref{fig:OmegaVsBVFreq}, we compare the Brunt-V{\"a}is{\"a}l{\"a} frequency ($N$) and the angular velocity($\Omega$) of the star, plotted on an equatorial line at $t=0$. We plot the Brunt-V{\"a}is{\"a}l{\"a} frequency only for the region inside the star $\varpi \lesssim 9$\,km. For the regions close to the surface, 9km~$\lesssim \varpi\lesssim 12$\,~km , the star is convectively unstable ($N^2 < 0$). In the convectively stable regions, the angular velocity is greater than the Brunt-V{\"a}is{\"a}l{\"a} frequency, $\Omega \gtrsim N$, violating the criterion for TSI instability. Thus, the scaling model for TSI instability would not have significant momentum transport for the remnant considered in this paper. 
 
As when comparing MRI models for this star, the result is sensitive to the configuration evolved, which although motivated by a merger simulation remains somewhat artificial.  It has no entropy gradient by construction, but $N$ for neutron stars (except perhaps very non-degenerate ones) is dominated by the $Y_e$ gradient.  The assumption of charged current weak nuclear equilibrium inside the star is a safe one, but the assumption of zero initial electron neutrino chemical potential is not very accurate and will affect the initial composition profile, and hence $N$.  For future post-merger simulations, we plan to use a simulation-motivated neutrino chemical potential profile.  Nevertheless, our finding for this star suggests that the $0<N<\Omega$ regime might in fact be important for some post-merger configurations.

\bibliography{reference}{}

\begin{thebibliography}{70}%
\makeatletter
\providecommand \@ifxundefined [1]{%
 \@ifx{#1\undefined}
}%
\providecommand \@ifnum [1]{%
 \ifnum #1\expandafter \@firstoftwo
 \else \expandafter \@secondoftwo
 \fi
}%
\providecommand \@ifx [1]{%
 \ifx #1\expandafter \@firstoftwo
 \else \expandafter \@secondoftwo
 \fi
}%
\providecommand \natexlab [1]{#1}%
\providecommand \enquote  [1]{``#1''}%
\providecommand \bibnamefont  [1]{#1}%
\providecommand \bibfnamefont [1]{#1}%
\providecommand \citenamefont [1]{#1}%
\providecommand \href@noop [0]{\@secondoftwo}%
\providecommand \href [0]{\begingroup \@sanitize@url \@href}%
\providecommand \@href[1]{\@@startlink{#1}\@@href}%
\providecommand \@@href[1]{\endgroup#1\@@endlink}%
\providecommand \@sanitize@url [0]{\catcode `\\12\catcode `\$12\catcode `\&12\catcode `\#12\catcode `\^12\catcode `\_12\catcode `\%12\relax}%
\providecommand \@@startlink[1]{}%
\providecommand \@@endlink[0]{}%
\providecommand \url  [0]{\begingroup\@sanitize@url \@url }%
\providecommand \@url [1]{\endgroup\@href {#1}{\urlprefix }}%
\providecommand \urlprefix  [0]{URL }%
\providecommand \Eprint [0]{\href }%
\providecommand \doibase [0]{https://doi.org/}%
\providecommand \selectlanguage [0]{\@gobble}%
\providecommand \bibinfo  [0]{\@secondoftwo}%
\providecommand \bibfield  [0]{\@secondoftwo}%
\providecommand \translation [1]{[#1]}%
\providecommand \BibitemOpen [0]{}%
\providecommand \bibitemStop [0]{}%
\providecommand \bibitemNoStop [0]{.\EOS\space}%
\providecommand \EOS [0]{\spacefactor3000\relax}%
\providecommand \BibitemShut  [1]{\csname bibitem#1\endcsname}%
\let\auto@bib@innerbib\@empty
\bibitem [{\citenamefont {Abbott}\ \emph {et~al.}(2017{\natexlab{a}})\citenamefont {Abbott} \emph {et~al.}}]{LIGOScientific:2017vwq}%
  \BibitemOpen
  \bibfield  {author} {\bibinfo {author} {\bibfnamefont {B.~P.}\ \bibnamefont {Abbott}} \emph {et~al.} (\bibinfo {collaboration} {LIGO Scientific, Virgo}),\ }\bibfield  {title} {\bibinfo {title} {{GW170817: Observation of Gravitational Waves from a Binary Neutron Star Inspiral}},\ }\href {https://doi.org/10.1103/PhysRevLett.119.161101} {\bibfield  {journal} {\bibinfo  {journal} {Phys. Rev. Lett.}\ }\textbf {\bibinfo {volume} {119}},\ \bibinfo {pages} {161101} (\bibinfo {year} {2017}{\natexlab{a}})},\ \Eprint {https://arxiv.org/abs/1710.05832} {arXiv:1710.05832 [gr-qc]} \BibitemShut {NoStop}%
\bibitem [{\citenamefont {Abbott}\ \emph {et~al.}(2020)\citenamefont {Abbott} \emph {et~al.}}]{LIGOScientific:2020aai}%
  \BibitemOpen
  \bibfield  {author} {\bibinfo {author} {\bibfnamefont {B.~P.}\ \bibnamefont {Abbott}} \emph {et~al.} (\bibinfo {collaboration} {LIGO Scientific, Virgo}),\ }\bibfield  {title} {\bibinfo {title} {{GW190425: Observation of a Compact Binary Coalescence with Total Mass $\sim 3.4 M_{\odot}$}},\ }\href {https://doi.org/10.3847/2041-8213/ab75f5} {\bibfield  {journal} {\bibinfo  {journal} {Astrophys. J. Lett.}\ }\textbf {\bibinfo {volume} {892}},\ \bibinfo {pages} {L3} (\bibinfo {year} {2020})},\ \Eprint {https://arxiv.org/abs/2001.01761} {arXiv:2001.01761 [astro-ph.HE]} \BibitemShut {NoStop}%
\bibitem [{\citenamefont {Abbott}\ \emph {et~al.}(2017{\natexlab{b}})\citenamefont {Abbott} \emph {et~al.}}]{LIGOScientific:2017ync}%
  \BibitemOpen
  \bibfield  {author} {\bibinfo {author} {\bibfnamefont {B.~P.}\ \bibnamefont {Abbott}} \emph {et~al.} (\bibinfo {collaboration} {LIGO Scientific, Virgo, Fermi GBM, INTEGRAL, IceCube, AstroSat Cadmium Zinc Telluride Imager Team, IPN, Insight-Hxmt, ANTARES, Swift, AGILE Team, 1M2H Team, Dark Energy Camera GW-EM, DES, DLT40, GRAWITA, Fermi-LAT, ATCA, ASKAP, Las Cumbres Observatory Group, OzGrav, DWF (Deeper Wider Faster Program), AST3, CAASTRO, VINROUGE, MASTER, J-GEM, GROWTH, JAGWAR, CaltechNRAO, TTU-NRAO, NuSTAR, Pan-STARRS, MAXI Team, TZAC Consortium, KU, Nordic Optical Telescope, ePESSTO, GROND, Texas Tech University, SALT Group, TOROS, BOOTES, MWA, CALET, IKI-GW Follow-up, H.E.S.S., LOFAR, LWA, HAWC, Pierre Auger, ALMA, Euro VLBI Team, Pi of Sky, Chandra Team at McGill University, DFN, ATLAS Telescopes, High Time Resolution Universe Survey, RIMAS, RATIR, SKA South Africa/MeerKAT}),\ }\bibfield  {title} {\bibinfo {title} {{Multi-messenger Observations of a Binary Neutron Star Merger}},\ }\href
  {https://doi.org/10.3847/2041-8213/aa91c9} {\bibfield  {journal} {\bibinfo  {journal} {Astrophys. J. Lett.}\ }\textbf {\bibinfo {volume} {848}},\ \bibinfo {pages} {L12} (\bibinfo {year} {2017}{\natexlab{b}})},\ \Eprint {https://arxiv.org/abs/1710.05833} {arXiv:1710.05833 [astro-ph.HE]} \BibitemShut {NoStop}%
\bibitem [{\citenamefont {{Baiotti}}\ and\ \citenamefont {{Rezzolla}}(2017)}]{Baiotti:2017}%
  \BibitemOpen
  \bibfield  {author} {\bibinfo {author} {\bibfnamefont {L.}~\bibnamefont {{Baiotti}}}\ and\ \bibinfo {author} {\bibfnamefont {L.}~\bibnamefont {{Rezzolla}}},\ }\bibfield  {title} {\bibinfo {title} {{Binary neutron star mergers: a review of Einstein{\textquoteright}s richest laboratory}},\ }\href {https://doi.org/10.1088/1361-6633/aa67bb} {\bibfield  {journal} {\bibinfo  {journal} {Reports on Progress in Physics}\ }\textbf {\bibinfo {volume} {80}},\ \bibinfo {eid} {096901} (\bibinfo {year} {2017})},\ \Eprint {https://arxiv.org/abs/1607.03540} {arXiv:1607.03540 [gr-qc]} \BibitemShut {NoStop}%
\bibitem [{\citenamefont {Burns}(2020)}]{Burns:2019}%
  \BibitemOpen
  \bibfield  {author} {\bibinfo {author} {\bibfnamefont {E.}~\bibnamefont {Burns}},\ }\bibfield  {title} {\bibinfo {title} {{Neutron Star Mergers and How to Study Them}},\ }\href {https://doi.org/10.1007/s41114-020-00028-7} {\bibfield  {journal} {\bibinfo  {journal} {Living Rev. Rel.}\ }\textbf {\bibinfo {volume} {23}},\ \bibinfo {pages} {4} (\bibinfo {year} {2020})},\ \Eprint {https://arxiv.org/abs/1909.06085} {arXiv:1909.06085 [astro-ph.HE]} \BibitemShut {NoStop}%
\bibitem [{\citenamefont {Faber}\ and\ \citenamefont {Rasio}(2012)}]{Faber:2012}%
  \BibitemOpen
  \bibfield  {author} {\bibinfo {author} {\bibfnamefont {J.~A.}\ \bibnamefont {Faber}}\ and\ \bibinfo {author} {\bibfnamefont {F.~A.}\ \bibnamefont {Rasio}},\ }\bibfield  {title} {\bibinfo {title} {{Binary Neutron Star Mergers}},\ }\href {https://doi.org/10.12942/lrr-2012-8} {\bibfield  {journal} {\bibinfo  {journal} {Living Rev. Rel.}\ }\textbf {\bibinfo {volume} {15}},\ \bibinfo {pages} {8} (\bibinfo {year} {2012})},\ \Eprint {https://arxiv.org/abs/1204.3858} {arXiv:1204.3858 [gr-qc]} \BibitemShut {NoStop}%
\bibitem [{\citenamefont {{Radice}}\ \emph {et~al.}(2020)\citenamefont {{Radice}}, \citenamefont {{Bernuzzi}},\ and\ \citenamefont {{Perego}}}]{RadiceReview:2020}%
  \BibitemOpen
  \bibfield  {author} {\bibinfo {author} {\bibfnamefont {D.}~\bibnamefont {{Radice}}}, \bibinfo {author} {\bibfnamefont {S.}~\bibnamefont {{Bernuzzi}}},\ and\ \bibinfo {author} {\bibfnamefont {A.}~\bibnamefont {{Perego}}},\ }\bibfield  {title} {\bibinfo {title} {{The Dynamics of Binary Neutron Star Mergers and GW170817}},\ }\href {https://doi.org/10.1146/annurev-nucl-013120-114541} {\bibfield  {journal} {\bibinfo  {journal} {Annual Review of Nuclear and Particle Science}\ }\textbf {\bibinfo {volume} {70}},\ \bibinfo {pages} {95} (\bibinfo {year} {2020})},\ \Eprint {https://arxiv.org/abs/2002.03863} {arXiv:2002.03863 [astro-ph.HE]} \BibitemShut {NoStop}%
\bibitem [{\citenamefont {Batchelor}(1969)}]{Batchelor:1969abc}%
  \BibitemOpen
  \bibfield  {author} {\bibinfo {author} {\bibfnamefont {G.~K.}\ \bibnamefont {Batchelor}},\ }\bibfield  {title} {\bibinfo {title} {Computation of the energy spectrum in homogeneous two-dimensional turbulence},\ }\href@noop {} {\bibfield  {journal} {\bibinfo  {journal} {Phys. Fluids}\ }\textbf {\bibinfo {volume} {12}},\ \bibinfo {pages} {II} (\bibinfo {year} {1969})}\BibitemShut {NoStop}%
\bibitem [{\citenamefont {Davidson}(2015)}]{Davidson:2015abc}%
  \BibitemOpen
  \bibfield  {author} {\bibinfo {author} {\bibfnamefont {P.~A.}\ \bibnamefont {Davidson}},\ }\href@noop {} {\emph {\bibinfo {title} {Turbulence: an introduction for scientists and engineers}}}\ (\bibinfo  {publisher} {Oxford university press},\ \bibinfo {year} {2015})\BibitemShut {NoStop}%
\bibitem [{\citenamefont {{Cowling}}(1933)}]{Cowling:1933abc}%
  \BibitemOpen
  \bibfield  {author} {\bibinfo {author} {\bibfnamefont {T.~G.}\ \bibnamefont {{Cowling}}},\ }\bibfield  {title} {\bibinfo {title} {{The magnetic field of sunspots}},\ }\href {https://doi.org/10.1093/mnras/94.1.39} {\bibfield  {journal} {\bibinfo  {journal} {Mon. Not. Roy. Astron. Soc.}\ }\textbf {\bibinfo {volume} {94}},\ \bibinfo {pages} {39} (\bibinfo {year} {1933})}\BibitemShut {NoStop}%
\bibitem [{\citenamefont {{Moffatt}}(1978)}]{Moffatt:1978abc}%
  \BibitemOpen
  \bibfield  {author} {\bibinfo {author} {\bibfnamefont {H.~K.}\ \bibnamefont {{Moffatt}}},\ }\href@noop {} {\emph {\bibinfo {title} {{Magnetic field generation in electrically conducting fluids}}}}\ (\bibinfo  {publisher} {Cambridge University Press},\ \bibinfo {address} {Cambridge},\ \bibinfo {year} {1978})\BibitemShut {NoStop}%
\bibitem [{\citenamefont {{Alcubierre}}\ \emph {et~al.}(2001)\citenamefont {{Alcubierre}}, \citenamefont {{Br{\"u}gmann}}, \citenamefont {{Holz}}, \citenamefont {{Takahashi}}, \citenamefont {{Brandt}}, \citenamefont {{Seidel}}, \citenamefont {{Thornburg}},\ and\ \citenamefont {{Ashtekar}}}]{Alcubierre:2001}%
  \BibitemOpen
  \bibfield  {author} {\bibinfo {author} {\bibfnamefont {M.}~\bibnamefont {{Alcubierre}}}, \bibinfo {author} {\bibfnamefont {B.}~\bibnamefont {{Br{\"u}gmann}}}, \bibinfo {author} {\bibfnamefont {D.}~\bibnamefont {{Holz}}}, \bibinfo {author} {\bibfnamefont {R.}~\bibnamefont {{Takahashi}}}, \bibinfo {author} {\bibfnamefont {S.}~\bibnamefont {{Brandt}}}, \bibinfo {author} {\bibfnamefont {E.}~\bibnamefont {{Seidel}}}, \bibinfo {author} {\bibfnamefont {J.}~\bibnamefont {{Thornburg}}},\ and\ \bibinfo {author} {\bibfnamefont {A.}~\bibnamefont {{Ashtekar}}},\ }\bibfield  {title} {\bibinfo {title} {{Symmetry Without Symmetry}},\ }\href {https://doi.org/10.1142/S0218271801000834} {\bibfield  {journal} {\bibinfo  {journal} {International Journal of Modern Physics D}\ }\textbf {\bibinfo {volume} {10}},\ \bibinfo {pages} {273} (\bibinfo {year} {2001})},\ \Eprint {https://arxiv.org/abs/gr-qc/9908012} {arXiv:gr-qc/9908012 [gr-qc]} \BibitemShut {NoStop}%
\bibitem [{\citenamefont {{Baumgarte}}\ \emph {et~al.}(2013)\citenamefont {{Baumgarte}}, \citenamefont {{Montero}}, \citenamefont {{Cordero-Carri{\'o}n}},\ and\ \citenamefont {{M{\"u}ller}}}]{BaumgarteEtAl:2013}%
  \BibitemOpen
  \bibfield  {author} {\bibinfo {author} {\bibfnamefont {T.~W.}\ \bibnamefont {{Baumgarte}}}, \bibinfo {author} {\bibfnamefont {P.~J.}\ \bibnamefont {{Montero}}}, \bibinfo {author} {\bibfnamefont {I.}~\bibnamefont {{Cordero-Carri{\'o}n}}},\ and\ \bibinfo {author} {\bibfnamefont {E.}~\bibnamefont {{M{\"u}ller}}},\ }\bibfield  {title} {\bibinfo {title} {{Numerical relativity in spherical polar coordinates: Evolution calculations with the BSSN formulation}},\ }\href {https://doi.org/10.1103/PhysRevD.87.044026} {\bibfield  {journal} {\bibinfo  {journal} {\prd}\ }\textbf {\bibinfo {volume} {87}},\ \bibinfo {eid} {044026} (\bibinfo {year} {2013})},\ \Eprint {https://arxiv.org/abs/1211.6632} {arXiv:1211.6632 [gr-qc]} \BibitemShut {NoStop}%
\bibitem [{\citenamefont {{Shibata}}(2000)}]{Shibata:2000}%
  \BibitemOpen
  \bibfield  {author} {\bibinfo {author} {\bibfnamefont {M.}~\bibnamefont {{Shibata}}},\ }\bibfield  {title} {\bibinfo {title} {{Axisymmetric Simulations of Rotating Stellar Collapse in Full General Relativity ---Criteria for Prompt Collapse to Black Holes---}},\ }\href {https://doi.org/10.1143/PTP.104.325} {\bibfield  {journal} {\bibinfo  {journal} {Progress of Theoretical Physics}\ }\textbf {\bibinfo {volume} {104}},\ \bibinfo {pages} {325} (\bibinfo {year} {2000})},\ \Eprint {https://arxiv.org/abs/gr-qc/0007049} {arXiv:gr-qc/0007049 [gr-qc]} \BibitemShut {NoStop}%
\bibitem [{\citenamefont {{Duez}}\ \emph {et~al.}(2004)\citenamefont {{Duez}}, \citenamefont {{Liu}}, \citenamefont {{Shapiro}},\ and\ \citenamefont {{Stephens}}}]{Matt:2004}%
  \BibitemOpen
  \bibfield  {author} {\bibinfo {author} {\bibfnamefont {M.~D.}\ \bibnamefont {{Duez}}}, \bibinfo {author} {\bibfnamefont {Y.~T.}\ \bibnamefont {{Liu}}}, \bibinfo {author} {\bibfnamefont {S.~L.}\ \bibnamefont {{Shapiro}}},\ and\ \bibinfo {author} {\bibfnamefont {B.~C.}\ \bibnamefont {{Stephens}}},\ }\bibfield  {title} {\bibinfo {title} {{General relativistic hydrodynamics with viscosity: Contraction, catastrophic collapse, and disk formation in hypermassive neutron stars}},\ }\href {https://doi.org/10.1103/PhysRevD.69.104030} {\bibfield  {journal} {\bibinfo  {journal} {\prd}\ }\textbf {\bibinfo {volume} {69}},\ \bibinfo {eid} {104030} (\bibinfo {year} {2004})},\ \Eprint {https://arxiv.org/abs/astro-ph/0402502} {arXiv:astro-ph/0402502 [astro-ph]} \BibitemShut {NoStop}%
\bibitem [{SpE()}]{SpEC}%
  \BibitemOpen
  \href@noop {} {\bibinfo {title} {{SpEC}: Spectral einstein code}},\ \bibinfo {howpublished} {\url{https://www.black-holes.org/code/SpEC.html}},\ \bibinfo {note} {[{A}ccessed May 6, 2025]}\BibitemShut {NoStop}%
\bibitem [{\citenamefont {{Lindblom}}\ \emph {et~al.}(2006)\citenamefont {{Lindblom}}, \citenamefont {{Scheel}}, \citenamefont {{Kidder}}, \citenamefont {{Owen}},\ and\ \citenamefont {{Rinne}}}]{Lindblom:2006}%
  \BibitemOpen
  \bibfield  {author} {\bibinfo {author} {\bibfnamefont {L.}~\bibnamefont {{Lindblom}}}, \bibinfo {author} {\bibfnamefont {M.~A.}\ \bibnamefont {{Scheel}}}, \bibinfo {author} {\bibfnamefont {L.~E.}\ \bibnamefont {{Kidder}}}, \bibinfo {author} {\bibfnamefont {R.}~\bibnamefont {{Owen}}},\ and\ \bibinfo {author} {\bibfnamefont {O.}~\bibnamefont {{Rinne}}},\ }\bibfield  {title} {\bibinfo {title} {{A new generalized harmonic evolution system}},\ }\href {https://doi.org/10.1088/0264-9381/23/16/S09} {\bibfield  {journal} {\bibinfo  {journal} {Classical and Quantum Gravity}\ }\textbf {\bibinfo {volume} {23}},\ \bibinfo {pages} {S447} (\bibinfo {year} {2006})},\ \Eprint {https://arxiv.org/abs/gr-qc/0512093} {arXiv:gr-qc/0512093 [gr-qc]} \BibitemShut {NoStop}%
\bibitem [{\citenamefont {Liu}\ \emph {et~al.}(1994)\citenamefont {Liu}, \citenamefont {Osher},\ and\ \citenamefont {Chan}}]{Liu:1994}%
  \BibitemOpen
  \bibfield  {author} {\bibinfo {author} {\bibfnamefont {X.-D.}\ \bibnamefont {Liu}}, \bibinfo {author} {\bibfnamefont {S.}~\bibnamefont {Osher}},\ and\ \bibinfo {author} {\bibfnamefont {T.}~\bibnamefont {Chan}},\ }\bibfield  {title} {\bibinfo {title} {Weighted essentially non-oscillatory schemes},\ }\href {https://doi.org/https://doi.org/10.1006/jcph.1994.1187} {\bibfield  {journal} {\bibinfo  {journal} {Journal of Computational Physics}\ }\textbf {\bibinfo {volume} {115}},\ \bibinfo {pages} {200} (\bibinfo {year} {1994})}\BibitemShut {NoStop}%
\bibitem [{\citenamefont {Jiang}\ and\ \citenamefont {Shu}(1996)}]{Jiang:1996}%
  \BibitemOpen
  \bibfield  {author} {\bibinfo {author} {\bibfnamefont {G.-S.}\ \bibnamefont {Jiang}}\ and\ \bibinfo {author} {\bibfnamefont {C.-W.}\ \bibnamefont {Shu}},\ }\bibfield  {title} {\bibinfo {title} {Efficient implementation of weighted eno schemes},\ }\href {https://doi.org/https://doi.org/10.1006/jcph.1996.0130} {\bibfield  {journal} {\bibinfo  {journal} {Journal of Computational Physics}\ }\textbf {\bibinfo {volume} {126}},\ \bibinfo {pages} {202} (\bibinfo {year} {1996})}\BibitemShut {NoStop}%
\bibitem [{\citenamefont {Harten}\ \emph {et~al.}(1983)\citenamefont {Harten}, \citenamefont {Lax},\ and\ \citenamefont {Leer}}]{HLL:1983}%
  \BibitemOpen
  \bibfield  {author} {\bibinfo {author} {\bibfnamefont {A.}~\bibnamefont {Harten}}, \bibinfo {author} {\bibfnamefont {P.~D.}\ \bibnamefont {Lax}},\ and\ \bibinfo {author} {\bibfnamefont {B.~v.}\ \bibnamefont {Leer}},\ }\bibfield  {title} {\bibinfo {title} {On upstream differencing and godunov-type schemes for hyperbolic conservation laws},\ }\href {https://doi.org/10.1137/1025002} {\bibfield  {journal} {\bibinfo  {journal} {SIAM Review}\ }\textbf {\bibinfo {volume} {25}},\ \bibinfo {pages} {35} (\bibinfo {year} {1983})},\ \Eprint {https://arxiv.org/abs/https://doi.org/10.1137/1025002} {https://doi.org/10.1137/1025002} \BibitemShut {NoStop}%
\bibitem [{\citenamefont {{Jesse}}\ \emph {et~al.}(2020)\citenamefont {{Jesse}}, \citenamefont {{Duez}}, \citenamefont {{Foucart}}, \citenamefont {{Haddadi}}, \citenamefont {{Knight}}, \citenamefont {{Cadenhead}}, \citenamefont {{H{\'e}bert}}, \citenamefont {{Kidder}}, \citenamefont {{Pfeiffer}},\ and\ \citenamefont {{Scheel}}}]{Jerred}%
  \BibitemOpen
  \bibfield  {author} {\bibinfo {author} {\bibfnamefont {J.}~\bibnamefont {{Jesse}}}, \bibinfo {author} {\bibfnamefont {M.~D.}\ \bibnamefont {{Duez}}}, \bibinfo {author} {\bibfnamefont {F.}~\bibnamefont {{Foucart}}}, \bibinfo {author} {\bibfnamefont {M.}~\bibnamefont {{Haddadi}}}, \bibinfo {author} {\bibfnamefont {A.~L.}\ \bibnamefont {{Knight}}}, \bibinfo {author} {\bibfnamefont {C.~L.}\ \bibnamefont {{Cadenhead}}}, \bibinfo {author} {\bibfnamefont {F.}~\bibnamefont {{H{\'e}bert}}}, \bibinfo {author} {\bibfnamefont {L.~E.}\ \bibnamefont {{Kidder}}}, \bibinfo {author} {\bibfnamefont {H.~P.}\ \bibnamefont {{Pfeiffer}}},\ and\ \bibinfo {author} {\bibfnamefont {M.~A.}\ \bibnamefont {{Scheel}}},\ }\bibfield  {title} {\bibinfo {title} {{Axisymmetric hydrodynamics in numerical relativity using a multipatch method}},\ }\href {https://doi.org/10.1088/1361-6382/abbc8b} {\bibfield  {journal} {\bibinfo  {journal} {Classical and Quantum Gravity}\ }\textbf {\bibinfo {volume} {37}},\ \bibinfo {eid} {235010} (\bibinfo
  {year} {2020})},\ \Eprint {https://arxiv.org/abs/2005.01848} {arXiv:2005.01848 [gr-qc]} \BibitemShut {NoStop}%
\bibitem [{\citenamefont {{Muhammed}}\ \emph {et~al.}(2024)\citenamefont {{Muhammed}}, \citenamefont {{Duez}}, \citenamefont {{Chawhan}}, \citenamefont {{Ghadiri}}, \citenamefont {{Buchman}}, \citenamefont {{Foucart}}, \citenamefont {{Cheong}}, \citenamefont {{Kidder}}, \citenamefont {{Pfeiffer}},\ and\ \citenamefont {{Scheel}}}]{Nishad:2024}%
  \BibitemOpen
  \bibfield  {author} {\bibinfo {author} {\bibfnamefont {N.}~\bibnamefont {{Muhammed}}}, \bibinfo {author} {\bibfnamefont {M.~D.}\ \bibnamefont {{Duez}}}, \bibinfo {author} {\bibfnamefont {P.}~\bibnamefont {{Chawhan}}}, \bibinfo {author} {\bibfnamefont {N.}~\bibnamefont {{Ghadiri}}}, \bibinfo {author} {\bibfnamefont {L.~T.}\ \bibnamefont {{Buchman}}}, \bibinfo {author} {\bibfnamefont {F.}~\bibnamefont {{Foucart}}}, \bibinfo {author} {\bibfnamefont {P.~C.-K.}\ \bibnamefont {{Cheong}}}, \bibinfo {author} {\bibfnamefont {L.~E.}\ \bibnamefont {{Kidder}}}, \bibinfo {author} {\bibfnamefont {H.~P.}\ \bibnamefont {{Pfeiffer}}},\ and\ \bibinfo {author} {\bibfnamefont {M.~A.}\ \bibnamefont {{Scheel}}},\ }\bibfield  {title} {\bibinfo {title} {{Stability of hypermassive neutron stars with realistic rotation and entropy profiles}},\ }\href {https://doi.org/10.1103/PhysRevD.110.124063} {\bibfield  {journal} {\bibinfo  {journal} {\prd}\ }\textbf {\bibinfo {volume} {110}},\ \bibinfo {eid} {124063} (\bibinfo {year} {2024})},\
  \Eprint {https://arxiv.org/abs/2403.05642} {arXiv:2403.05642 [gr-qc]} \BibitemShut {NoStop}%
\bibitem [{\citenamefont {Pretorius}(2005)}]{Pretorius:2005}%
  \BibitemOpen
  \bibfield  {author} {\bibinfo {author} {\bibfnamefont {F.}~\bibnamefont {Pretorius}},\ }\bibfield  {title} {\bibinfo {title} {{Numerical relativity using a generalized harmonic decomposition}},\ }\href {https://doi.org/10.1088/0264-9381/22/2/014} {\bibfield  {journal} {\bibinfo  {journal} {Class. Quant. Grav.}\ }\textbf {\bibinfo {volume} {22}},\ \bibinfo {pages} {425} (\bibinfo {year} {2005})},\ \Eprint {https://arxiv.org/abs/gr-qc/0407110} {arXiv:gr-qc/0407110} \BibitemShut {NoStop}%
\bibitem [{\citenamefont {Hilditch}\ \emph {et~al.}(2016)\citenamefont {Hilditch}, \citenamefont {Weyhausen},\ and\ \citenamefont {Br\"ugmann}}]{Hilditch:2015}%
  \BibitemOpen
  \bibfield  {author} {\bibinfo {author} {\bibfnamefont {D.}~\bibnamefont {Hilditch}}, \bibinfo {author} {\bibfnamefont {A.}~\bibnamefont {Weyhausen}},\ and\ \bibinfo {author} {\bibfnamefont {B.}~\bibnamefont {Br\"ugmann}},\ }\bibfield  {title} {\bibinfo {title} {{Pseudospectral method for gravitational wave collapse}},\ }\href {https://doi.org/10.1103/PhysRevD.93.063006} {\bibfield  {journal} {\bibinfo  {journal} {Phys. Rev. D}\ }\textbf {\bibinfo {volume} {93}},\ \bibinfo {pages} {063006} (\bibinfo {year} {2016})},\ \Eprint {https://arxiv.org/abs/1504.04732} {arXiv:1504.04732 [gr-qc]} \BibitemShut {NoStop}%
\bibitem [{\citenamefont {Foucart}\ \emph {et~al.}(2015)\citenamefont {Foucart}, \citenamefont {O'Connor}, \citenamefont {Roberts}, \citenamefont {Duez}, \citenamefont {Haas}, \citenamefont {Kidder}, \citenamefont {Ott}, \citenamefont {Pfeiffer}, \citenamefont {Scheel},\ and\ \citenamefont {Szilagyi}}]{Foucart:2015vpa}%
  \BibitemOpen
  \bibfield  {author} {\bibinfo {author} {\bibfnamefont {F.}~\bibnamefont {Foucart}}, \bibinfo {author} {\bibfnamefont {E.}~\bibnamefont {O'Connor}}, \bibinfo {author} {\bibfnamefont {L.}~\bibnamefont {Roberts}}, \bibinfo {author} {\bibfnamefont {M.~D.}\ \bibnamefont {Duez}}, \bibinfo {author} {\bibfnamefont {R.}~\bibnamefont {Haas}}, \bibinfo {author} {\bibfnamefont {L.~E.}\ \bibnamefont {Kidder}}, \bibinfo {author} {\bibfnamefont {C.~D.}\ \bibnamefont {Ott}}, \bibinfo {author} {\bibfnamefont {H.~P.}\ \bibnamefont {Pfeiffer}}, \bibinfo {author} {\bibfnamefont {M.~A.}\ \bibnamefont {Scheel}},\ and\ \bibinfo {author} {\bibfnamefont {B.}~\bibnamefont {Szilagyi}},\ }\bibfield  {title} {\bibinfo {title} {{Post-merger evolution of a neutron star-black hole binary with neutrino transport}},\ }\href {https://doi.org/10.1103/PhysRevD.91.124021} {\bibfield  {journal} {\bibinfo  {journal} {Phys. Rev. D}\ }\textbf {\bibinfo {volume} {91}},\ \bibinfo {pages} {124021} (\bibinfo {year} {2015})},\ \Eprint
  {https://arxiv.org/abs/1502.04146} {arXiv:1502.04146 [astro-ph.HE]} \BibitemShut {NoStop}%
\bibitem [{\citenamefont {Foucart}\ \emph {et~al.}(2016)\citenamefont {Foucart}, \citenamefont {O'Connor}, \citenamefont {Roberts}, \citenamefont {Kidder}, \citenamefont {Pfeiffer},\ and\ \citenamefont {Scheel}}]{Foucart:2016rxm}%
  \BibitemOpen
  \bibfield  {author} {\bibinfo {author} {\bibfnamefont {F.}~\bibnamefont {Foucart}}, \bibinfo {author} {\bibfnamefont {E.}~\bibnamefont {O'Connor}}, \bibinfo {author} {\bibfnamefont {L.}~\bibnamefont {Roberts}}, \bibinfo {author} {\bibfnamefont {L.~E.}\ \bibnamefont {Kidder}}, \bibinfo {author} {\bibfnamefont {H.~P.}\ \bibnamefont {Pfeiffer}},\ and\ \bibinfo {author} {\bibfnamefont {M.~A.}\ \bibnamefont {Scheel}},\ }\bibfield  {title} {\bibinfo {title} {{Impact of an improved neutrino energy estimate on outflows in neutron star merger simulations}},\ }\href {https://doi.org/10.1103/PhysRevD.94.123016} {\bibfield  {journal} {\bibinfo  {journal} {Phys. Rev. D}\ }\textbf {\bibinfo {volume} {94}},\ \bibinfo {pages} {123016} (\bibinfo {year} {2016})},\ \Eprint {https://arxiv.org/abs/1607.07450} {arXiv:1607.07450 [astro-ph.HE]} \BibitemShut {NoStop}%
\bibitem [{\citenamefont {{Foucart}}\ \emph {et~al.}(2024)\citenamefont {{Foucart}}, \citenamefont {{Cheong}}, \citenamefont {{Duez}}, \citenamefont {{Kidder}}, \citenamefont {{Pfeiffer}},\ and\ \citenamefont {{Scheel}}}]{FrancoisNeutrinoTransport:2024}%
  \BibitemOpen
  \bibfield  {author} {\bibinfo {author} {\bibfnamefont {F.}~\bibnamefont {{Foucart}}}, \bibinfo {author} {\bibfnamefont {P.~C.-K.}\ \bibnamefont {{Cheong}}}, \bibinfo {author} {\bibfnamefont {M.~D.}\ \bibnamefont {{Duez}}}, \bibinfo {author} {\bibfnamefont {L.~E.}\ \bibnamefont {{Kidder}}}, \bibinfo {author} {\bibfnamefont {H.~P.}\ \bibnamefont {{Pfeiffer}}},\ and\ \bibinfo {author} {\bibfnamefont {M.~A.}\ \bibnamefont {{Scheel}}},\ }\bibfield  {title} {\bibinfo {title} {{Robustness of neutron star merger simulations to changes in neutrino transport and neutrino-matter interactions}},\ }\href {https://doi.org/10.1103/PhysRevD.110.083028} {\bibfield  {journal} {\bibinfo  {journal} {\prd}\ }\textbf {\bibinfo {volume} {110}},\ \bibinfo {eid} {083028} (\bibinfo {year} {2024})},\ \Eprint {https://arxiv.org/abs/2407.15989} {arXiv:2407.15989 [astro-ph.HE]} \BibitemShut {NoStop}%
\bibitem [{\citenamefont {{Radice}}\ \emph {et~al.}(2022)\citenamefont {{Radice}}, \citenamefont {{Bernuzzi}}, \citenamefont {{Perego}},\ and\ \citenamefont {{Haas}}}]{RadiceM1:2021}%
  \BibitemOpen
  \bibfield  {author} {\bibinfo {author} {\bibfnamefont {D.}~\bibnamefont {{Radice}}}, \bibinfo {author} {\bibfnamefont {S.}~\bibnamefont {{Bernuzzi}}}, \bibinfo {author} {\bibfnamefont {A.}~\bibnamefont {{Perego}}},\ and\ \bibinfo {author} {\bibfnamefont {R.}~\bibnamefont {{Haas}}},\ }\bibfield  {title} {\bibinfo {title} {{A new moment-based general-relativistic neutrino-radiation transport code: Methods and first applications to neutron star mergers}},\ }\href {https://doi.org/10.1093/mnras/stac589} {\bibfield  {journal} {\bibinfo  {journal} {\mnras}\ }\textbf {\bibinfo {volume} {512}},\ \bibinfo {pages} {1499} (\bibinfo {year} {2022})},\ \Eprint {https://arxiv.org/abs/2111.14858} {arXiv:2111.14858 [astro-ph.HE]} \BibitemShut {NoStop}%
\bibitem [{\citenamefont {Typel}\ \emph {et~al.}(2010)\citenamefont {Typel}, \citenamefont {R\"opke}, \citenamefont {Kl\"ahn}, \citenamefont {Blaschke},\ and\ \citenamefont {Wolter}}]{DD2}%
  \BibitemOpen
  \bibfield  {author} {\bibinfo {author} {\bibfnamefont {S.}~\bibnamefont {Typel}}, \bibinfo {author} {\bibfnamefont {G.}~\bibnamefont {R\"opke}}, \bibinfo {author} {\bibfnamefont {T.}~\bibnamefont {Kl\"ahn}}, \bibinfo {author} {\bibfnamefont {D.}~\bibnamefont {Blaschke}},\ and\ \bibinfo {author} {\bibfnamefont {H.~H.}\ \bibnamefont {Wolter}},\ }\bibfield  {title} {\bibinfo {title} {Composition and thermodynamics of nuclear matter with light clusters},\ }\href {https://doi.org/10.1103/PhysRevC.81.015803} {\bibfield  {journal} {\bibinfo  {journal} {Phys. Rev. C}\ }\textbf {\bibinfo {volume} {81}},\ \bibinfo {pages} {015803} (\bibinfo {year} {2010})}\BibitemShut {NoStop}%
\bibitem [{\citenamefont {{Vincent}}\ \emph {et~al.}(2020)\citenamefont {{Vincent}}, \citenamefont {{Foucart}}, \citenamefont {{Duez}}, \citenamefont {{Haas}}, \citenamefont {{Kidder}}, \citenamefont {{Pfeiffer}},\ and\ \citenamefont {{Scheel}}}]{Vincent:2020}%
  \BibitemOpen
  \bibfield  {author} {\bibinfo {author} {\bibfnamefont {T.}~\bibnamefont {{Vincent}}}, \bibinfo {author} {\bibfnamefont {F.}~\bibnamefont {{Foucart}}}, \bibinfo {author} {\bibfnamefont {M.~D.}\ \bibnamefont {{Duez}}}, \bibinfo {author} {\bibfnamefont {R.}~\bibnamefont {{Haas}}}, \bibinfo {author} {\bibfnamefont {L.~E.}\ \bibnamefont {{Kidder}}}, \bibinfo {author} {\bibfnamefont {H.~P.}\ \bibnamefont {{Pfeiffer}}},\ and\ \bibinfo {author} {\bibfnamefont {M.~A.}\ \bibnamefont {{Scheel}}},\ }\bibfield  {title} {\bibinfo {title} {{Unequal mass binary neutron star simulations with neutrino transport: Ejecta and neutrino emission}},\ }\href {https://doi.org/10.1103/PhysRevD.101.044053} {\bibfield  {journal} {\bibinfo  {journal} {\prd}\ }\textbf {\bibinfo {volume} {101}},\ \bibinfo {eid} {044053} (\bibinfo {year} {2020})},\ \Eprint {https://arxiv.org/abs/1908.00655} {arXiv:1908.00655 [gr-qc]} \BibitemShut {NoStop}%
\bibitem [{\citenamefont {{Cook}}\ \emph {et~al.}(1992)\citenamefont {{Cook}}, \citenamefont {{Shapiro}},\ and\ \citenamefont {{Teukolsky}}}]{Cook:1992}%
  \BibitemOpen
  \bibfield  {author} {\bibinfo {author} {\bibfnamefont {G.~B.}\ \bibnamefont {{Cook}}}, \bibinfo {author} {\bibfnamefont {S.~L.}\ \bibnamefont {{Shapiro}}},\ and\ \bibinfo {author} {\bibfnamefont {S.~A.}\ \bibnamefont {{Teukolsky}}},\ }\bibfield  {title} {\bibinfo {title} {{Spin-up of a Rapidly Rotating Star by Angular Momentum Loss: Effects of General Relativity}},\ }\href {https://doi.org/10.1086/171849} {\bibfield  {journal} {\bibinfo  {journal} {\apj}\ }\textbf {\bibinfo {volume} {398}},\ \bibinfo {pages} {203} (\bibinfo {year} {1992})}\BibitemShut {NoStop}%
\bibitem [{\citenamefont {{Cook}}\ \emph {et~al.}(1994)\citenamefont {{Cook}}, \citenamefont {{Shapiro}},\ and\ \citenamefont {{Teukolsky}}}]{Cook:1994}%
  \BibitemOpen
  \bibfield  {author} {\bibinfo {author} {\bibfnamefont {G.~B.}\ \bibnamefont {{Cook}}}, \bibinfo {author} {\bibfnamefont {S.~L.}\ \bibnamefont {{Shapiro}}},\ and\ \bibinfo {author} {\bibfnamefont {S.~A.}\ \bibnamefont {{Teukolsky}}},\ }\bibfield  {title} {\bibinfo {title} {{Rapidly Rotating Polytropes in General Relativity}},\ }\href {https://doi.org/10.1086/173721} {\bibfield  {journal} {\bibinfo  {journal} {\apj}\ }\textbf {\bibinfo {volume} {422}},\ \bibinfo {pages} {227} (\bibinfo {year} {1994})}\BibitemShut {NoStop}%
\bibitem [{\citenamefont {Kastaun}\ and\ \citenamefont {Galeazzi}(2015)}]{Kastaun:2014fna}%
  \BibitemOpen
  \bibfield  {author} {\bibinfo {author} {\bibfnamefont {W.}~\bibnamefont {Kastaun}}\ and\ \bibinfo {author} {\bibfnamefont {F.}~\bibnamefont {Galeazzi}},\ }\bibfield  {title} {\bibinfo {title} {{Properties of hypermassive neutron stars formed in mergers of spinning binaries}},\ }\href {https://doi.org/10.1103/PhysRevD.91.064027} {\bibfield  {journal} {\bibinfo  {journal} {Phys. Rev.}\ }\textbf {\bibinfo {volume} {D91}},\ \bibinfo {pages} {064027} (\bibinfo {year} {2015})},\ \Eprint {https://arxiv.org/abs/1411.7975} {arXiv:1411.7975 [gr-qc]} \BibitemShut {NoStop}%
\bibitem [{\citenamefont {Hanauske}\ \emph {et~al.}(2017)\citenamefont {Hanauske}, \citenamefont {Takami}, \citenamefont {Bovard}, \citenamefont {Rezzolla}, \citenamefont {Font}, \citenamefont {Galeazzi},\ and\ \citenamefont {St\"ocker}}]{Hanauske:2016gia}%
  \BibitemOpen
  \bibfield  {author} {\bibinfo {author} {\bibfnamefont {M.}~\bibnamefont {Hanauske}}, \bibinfo {author} {\bibfnamefont {K.}~\bibnamefont {Takami}}, \bibinfo {author} {\bibfnamefont {L.}~\bibnamefont {Bovard}}, \bibinfo {author} {\bibfnamefont {L.}~\bibnamefont {Rezzolla}}, \bibinfo {author} {\bibfnamefont {J.~A.}\ \bibnamefont {Font}}, \bibinfo {author} {\bibfnamefont {F.}~\bibnamefont {Galeazzi}},\ and\ \bibinfo {author} {\bibfnamefont {H.}~\bibnamefont {St\"ocker}},\ }\bibfield  {title} {\bibinfo {title} {{Rotational properties of hypermassive neutron stars from binary mergers}},\ }\href {https://doi.org/10.1103/PhysRevD.96.043004} {\bibfield  {journal} {\bibinfo  {journal} {Phys. Rev. D}\ }\textbf {\bibinfo {volume} {96}},\ \bibinfo {pages} {043004} (\bibinfo {year} {2017})},\ \Eprint {https://arxiv.org/abs/1611.07152} {arXiv:1611.07152 [gr-qc]} \BibitemShut {NoStop}%
\bibitem [{\citenamefont {De~Pietri}\ \emph {et~al.}(2020)\citenamefont {De~Pietri}, \citenamefont {Feo}, \citenamefont {Font}, \citenamefont {L{\"o}ffler}, \citenamefont {Pasquali},\ and\ \citenamefont {Stergioulas}}]{de2020numerical}%
  \BibitemOpen
  \bibfield  {author} {\bibinfo {author} {\bibfnamefont {R.}~\bibnamefont {De~Pietri}}, \bibinfo {author} {\bibfnamefont {A.}~\bibnamefont {Feo}}, \bibinfo {author} {\bibfnamefont {J.~A.}\ \bibnamefont {Font}}, \bibinfo {author} {\bibfnamefont {F.}~\bibnamefont {L{\"o}ffler}}, \bibinfo {author} {\bibfnamefont {M.}~\bibnamefont {Pasquali}},\ and\ \bibinfo {author} {\bibfnamefont {N.}~\bibnamefont {Stergioulas}},\ }\bibfield  {title} {\bibinfo {title} {Numerical-relativity simulations of long-lived remnants of binary neutron star mergers},\ }\href@noop {} {\bibfield  {journal} {\bibinfo  {journal} {Physical Review D}\ }\textbf {\bibinfo {volume} {101}},\ \bibinfo {pages} {064052} (\bibinfo {year} {2020})}\BibitemShut {NoStop}%
\bibitem [{\citenamefont {Iosif}\ and\ \citenamefont {Stergioulas}(2021)}]{iosif2021equilibrium}%
  \BibitemOpen
  \bibfield  {author} {\bibinfo {author} {\bibfnamefont {P.}~\bibnamefont {Iosif}}\ and\ \bibinfo {author} {\bibfnamefont {N.}~\bibnamefont {Stergioulas}},\ }\bibfield  {title} {\bibinfo {title} {Equilibrium sequences of differentially rotating stars with post-merger-like rotational profiles},\ }\href@noop {} {\bibfield  {journal} {\bibinfo  {journal} {Monthly Notices of the Royal Astronomical Society}\ }\textbf {\bibinfo {volume} {503}},\ \bibinfo {pages} {850} (\bibinfo {year} {2021})}\BibitemShut {NoStop}%
\bibitem [{\citenamefont {Iosif}\ and\ \citenamefont {Stergioulas}(2022)}]{iosif2022models}%
  \BibitemOpen
  \bibfield  {author} {\bibinfo {author} {\bibfnamefont {P.}~\bibnamefont {Iosif}}\ and\ \bibinfo {author} {\bibfnamefont {N.}~\bibnamefont {Stergioulas}},\ }\bibfield  {title} {\bibinfo {title} {Models of binary neutron star remnants with tabulated equations of state},\ }\href@noop {} {\bibfield  {journal} {\bibinfo  {journal} {Monthly Notices of the Royal Astronomical Society}\ }\textbf {\bibinfo {volume} {510}},\ \bibinfo {pages} {2948} (\bibinfo {year} {2022})}\BibitemShut {NoStop}%
\bibitem [{\citenamefont {{Cassing}}\ and\ \citenamefont {{Rezzolla}}(2024)}]{cassing2024realistic}%
  \BibitemOpen
  \bibfield  {author} {\bibinfo {author} {\bibfnamefont {M.}~\bibnamefont {{Cassing}}}\ and\ \bibinfo {author} {\bibfnamefont {L.}~\bibnamefont {{Rezzolla}}},\ }\bibfield  {title} {\bibinfo {title} {{Realistic models of general-relativistic differentially rotating stars}},\ }\href {https://doi.org/10.1093/mnras/stae1527} {\bibfield  {journal} {\bibinfo  {journal} {\mnras}\ }\textbf {\bibinfo {volume} {532}},\ \bibinfo {pages} {945} (\bibinfo {year} {2024})},\ \Eprint {https://arxiv.org/abs/2405.06609} {arXiv:2405.06609 [gr-qc]} \BibitemShut {NoStop}%
\bibitem [{\citenamefont {Staykov}\ \emph {et~al.}(2023)\citenamefont {Staykov}, \citenamefont {Doneva}, \citenamefont {Heisenberg}, \citenamefont {Stergioulas},\ and\ \citenamefont {Yazadjiev}}]{staykov2023differentially}%
  \BibitemOpen
  \bibfield  {author} {\bibinfo {author} {\bibfnamefont {K.~V.}\ \bibnamefont {Staykov}}, \bibinfo {author} {\bibfnamefont {D.~D.}\ \bibnamefont {Doneva}}, \bibinfo {author} {\bibfnamefont {L.}~\bibnamefont {Heisenberg}}, \bibinfo {author} {\bibfnamefont {N.}~\bibnamefont {Stergioulas}},\ and\ \bibinfo {author} {\bibfnamefont {S.~S.}\ \bibnamefont {Yazadjiev}},\ }\bibfield  {title} {\bibinfo {title} {Differentially rotating scalarized neutron stars with realistic postmerger profiles},\ }\href@noop {} {\bibfield  {journal} {\bibinfo  {journal} {Physical Review D}\ }\textbf {\bibinfo {volume} {108}},\ \bibinfo {pages} {024058} (\bibinfo {year} {2023})}\BibitemShut {NoStop}%
\bibitem [{\citenamefont {Uryu}\ \emph {et~al.}(2017)\citenamefont {Uryu}, \citenamefont {Tsokaros}, \citenamefont {Baiotti}, \citenamefont {Galeazzi}, \citenamefont {Taniguchi},\ and\ \citenamefont {Yoshida}}]{Uryu:2017obi}%
  \BibitemOpen
  \bibfield  {author} {\bibinfo {author} {\bibfnamefont {K.}~\bibnamefont {Uryu}}, \bibinfo {author} {\bibfnamefont {A.}~\bibnamefont {Tsokaros}}, \bibinfo {author} {\bibfnamefont {L.}~\bibnamefont {Baiotti}}, \bibinfo {author} {\bibfnamefont {F.}~\bibnamefont {Galeazzi}}, \bibinfo {author} {\bibfnamefont {K.}~\bibnamefont {Taniguchi}},\ and\ \bibinfo {author} {\bibfnamefont {S.}~\bibnamefont {Yoshida}},\ }\bibfield  {title} {\bibinfo {title} {{Modeling differential rotations of compact stars in equilibriums}},\ }\href {https://doi.org/10.1103/PhysRevD.96.103011} {\bibfield  {journal} {\bibinfo  {journal} {Phys. Rev. D}\ }\textbf {\bibinfo {volume} {96}},\ \bibinfo {pages} {103011} (\bibinfo {year} {2017})},\ \Eprint {https://arxiv.org/abs/1709.02643} {arXiv:1709.02643 [astro-ph.HE]} \BibitemShut {NoStop}%
\bibitem [{\citenamefont {Shibata}(2003)}]{Shibata:2003}%
  \BibitemOpen
  \bibfield  {author} {\bibinfo {author} {\bibfnamefont {M.}~\bibnamefont {Shibata}},\ }\bibfield  {title} {\bibinfo {title} {Axisymmetric general relativistic hydrodynamics: Long-term evolution of neutron stars and stellar collapse to neutron stars and black holes},\ }\href {https://doi.org/10.1103/PhysRevD.67.024033} {\bibfield  {journal} {\bibinfo  {journal} {Phys. Rev. D}\ }\textbf {\bibinfo {volume} {67}},\ \bibinfo {pages} {024033} (\bibinfo {year} {2003})}\BibitemShut {NoStop}%
\bibitem [{\citenamefont {{Montero}}\ \emph {et~al.}(2014)\citenamefont {{Montero}}, \citenamefont {{Baumgarte}},\ and\ \citenamefont {{M{\"u}ller}}}]{MonteroEtAl:2014}%
  \BibitemOpen
  \bibfield  {author} {\bibinfo {author} {\bibfnamefont {P.~J.}\ \bibnamefont {{Montero}}}, \bibinfo {author} {\bibfnamefont {T.~W.}\ \bibnamefont {{Baumgarte}}},\ and\ \bibinfo {author} {\bibfnamefont {E.}~\bibnamefont {{M{\"u}ller}}},\ }\bibfield  {title} {\bibinfo {title} {{General relativistic hydrodynamics in curvilinear coordinates}},\ }\href {https://doi.org/10.1103/PhysRevD.89.084043} {\bibfield  {journal} {\bibinfo  {journal} {\prd}\ }\textbf {\bibinfo {volume} {89}},\ \bibinfo {eid} {084043} (\bibinfo {year} {2014})},\ \Eprint {https://arxiv.org/abs/1309.7808} {arXiv:1309.7808 [gr-qc]} \BibitemShut {NoStop}%
\bibitem [{\citenamefont {Baumgarte}\ \emph {et~al.}(2015)\citenamefont {Baumgarte}, \citenamefont {Montero},\ and\ \citenamefont {M\"uller}}]{BaumgarteMoneteroMüller:2015}%
  \BibitemOpen
  \bibfield  {author} {\bibinfo {author} {\bibfnamefont {T.~W.}\ \bibnamefont {Baumgarte}}, \bibinfo {author} {\bibfnamefont {P.~J.}\ \bibnamefont {Montero}},\ and\ \bibinfo {author} {\bibfnamefont {E.}~\bibnamefont {M\"uller}},\ }\bibfield  {title} {\bibinfo {title} {Numerical relativity in spherical polar coordinates: Off-center simulations},\ }\href {https://doi.org/10.1103/PhysRevD.91.064035} {\bibfield  {journal} {\bibinfo  {journal} {Phys. Rev. D}\ }\textbf {\bibinfo {volume} {91}},\ \bibinfo {pages} {064035} (\bibinfo {year} {2015})}\BibitemShut {NoStop}%
\bibitem [{\citenamefont {{Mewes}}\ \emph {et~al.}(2020)\citenamefont {{Mewes}}, \citenamefont {{Zlochower}}, \citenamefont {{Campanelli}}, \citenamefont {{Baumgarte}}, \citenamefont {{Etienne}}, \citenamefont {{Armengol}},\ and\ \citenamefont {{Cipolletta}}}]{Mewes:2020}%
  \BibitemOpen
  \bibfield  {author} {\bibinfo {author} {\bibfnamefont {V.}~\bibnamefont {{Mewes}}}, \bibinfo {author} {\bibfnamefont {Y.}~\bibnamefont {{Zlochower}}}, \bibinfo {author} {\bibfnamefont {M.}~\bibnamefont {{Campanelli}}}, \bibinfo {author} {\bibfnamefont {T.~W.}\ \bibnamefont {{Baumgarte}}}, \bibinfo {author} {\bibfnamefont {Z.~B.}\ \bibnamefont {{Etienne}}}, \bibinfo {author} {\bibfnamefont {F.~G.~L.}\ \bibnamefont {{Armengol}}},\ and\ \bibinfo {author} {\bibfnamefont {F.}~\bibnamefont {{Cipolletta}}},\ }\bibfield  {title} {\bibinfo {title} {{Numerical relativity in spherical coordinates: A new dynamical spacetime and general relativistic MHD evolution framework for the Einstein Toolkit}},\ }\href {https://doi.org/10.1103/PhysRevD.101.104007} {\bibfield  {journal} {\bibinfo  {journal} {\prd}\ }\textbf {\bibinfo {volume} {101}},\ \bibinfo {eid} {104007} (\bibinfo {year} {2020})},\ \Eprint {https://arxiv.org/abs/2002.06225} {arXiv:2002.06225 [gr-qc]} \BibitemShut {NoStop}%
\bibitem [{\citenamefont {{Jacques}}\ \emph {et~al.}(2024)\citenamefont {{Jacques}}, \citenamefont {{Cupp}}, \citenamefont {{Werneck}}, \citenamefont {{Tootle}}, \citenamefont {{Babiuc Hamilton}},\ and\ \citenamefont {{Etienne}}}]{Terrence:2024}%
  \BibitemOpen
  \bibfield  {author} {\bibinfo {author} {\bibfnamefont {T.~P.}\ \bibnamefont {{Jacques}}}, \bibinfo {author} {\bibfnamefont {S.}~\bibnamefont {{Cupp}}}, \bibinfo {author} {\bibfnamefont {L.~R.}\ \bibnamefont {{Werneck}}}, \bibinfo {author} {\bibfnamefont {S.~D.}\ \bibnamefont {{Tootle}}}, \bibinfo {author} {\bibfnamefont {M.~C.}\ \bibnamefont {{Babiuc Hamilton}}},\ and\ \bibinfo {author} {\bibfnamefont {Z.~B.}\ \bibnamefont {{Etienne}}},\ }\bibfield  {title} {\bibinfo {title} {{GRoovy: A General Relativistic Hydrodynamics Code for Dynamical Spacetimes with Curvilinear Coordinates, Tabulated Equations of State, and Neutrino Physics}},\ }\href {https://doi.org/10.48550/arXiv.2412.03659} {\bibfield  {journal} {\bibinfo  {journal} {arXiv e-prints}\ ,\ \bibinfo {eid} {arXiv:2412.03659}} (\bibinfo {year} {2024})},\ \Eprint {https://arxiv.org/abs/2412.03659} {arXiv:2412.03659 [gr-qc]} \BibitemShut {NoStop}%
\bibitem [{\citenamefont {{Lam}}\ and\ \citenamefont {{Shibata}}(2025)}]{Lam:2025}%
  \BibitemOpen
  \bibfield  {author} {\bibinfo {author} {\bibfnamefont {A.~T.-L.}\ \bibnamefont {{Lam}}}\ and\ \bibinfo {author} {\bibfnamefont {M.}~\bibnamefont {{Shibata}}},\ }\bibfield  {title} {\bibinfo {title} {{New axisymmetric general relativistic hydrodynamics code with fixed mesh refinement}},\ }\href {https://doi.org/10.1103/PhysRevD.111.103039} {\bibfield  {journal} {\bibinfo  {journal} {\prd}\ }\textbf {\bibinfo {volume} {111}},\ \bibinfo {eid} {103039} (\bibinfo {year} {2025})},\ \Eprint {https://arxiv.org/abs/2502.03223} {arXiv:2502.03223 [astro-ph.HE]} \BibitemShut {NoStop}%
\bibitem [{\citenamefont {{Cheong}}\ \emph {et~al.}(2021)\citenamefont {{Cheong}}, \citenamefont {{Lam}}, \citenamefont {{Ng}},\ and\ \citenamefont {{Li}}}]{Patrick:Gmunu}%
  \BibitemOpen
  \bibfield  {author} {\bibinfo {author} {\bibfnamefont {P.~C.-K.}\ \bibnamefont {{Cheong}}}, \bibinfo {author} {\bibfnamefont {A.~T.-L.}\ \bibnamefont {{Lam}}}, \bibinfo {author} {\bibfnamefont {H.~H.-Y.}\ \bibnamefont {{Ng}}},\ and\ \bibinfo {author} {\bibfnamefont {T.~G.~F.}\ \bibnamefont {{Li}}},\ }\bibfield  {title} {\bibinfo {title} {{Gmunu: paralleled, grid-adaptive, general-relativistic magnetohydrodynamics in curvilinear geometries in dynamical space-times}},\ }\href {https://doi.org/10.1093/mnras/stab2606} {\bibfield  {journal} {\bibinfo  {journal} {\mnras}\ }\textbf {\bibinfo {volume} {508}},\ \bibinfo {pages} {2279} (\bibinfo {year} {2021})},\ \Eprint {https://arxiv.org/abs/2012.07322} {arXiv:2012.07322 [astro-ph.IM]} \BibitemShut {NoStop}%
\bibitem [{\citenamefont {{Cheong}}\ \emph {et~al.}(2024)\citenamefont {{Cheong}}, \citenamefont {{Muhammed}}, \citenamefont {{Chawhan}}, \citenamefont {{Duez}}, \citenamefont {{Foucart}}, \citenamefont {{Kidder}}, \citenamefont {{Pfeiffer}},\ and\ \citenamefont {{Scheel}}}]{Patrick:2024}%
  \BibitemOpen
  \bibfield  {author} {\bibinfo {author} {\bibfnamefont {P.~C.-K.}\ \bibnamefont {{Cheong}}}, \bibinfo {author} {\bibfnamefont {N.}~\bibnamefont {{Muhammed}}}, \bibinfo {author} {\bibfnamefont {P.}~\bibnamefont {{Chawhan}}}, \bibinfo {author} {\bibfnamefont {M.~D.}\ \bibnamefont {{Duez}}}, \bibinfo {author} {\bibfnamefont {F.}~\bibnamefont {{Foucart}}}, \bibinfo {author} {\bibfnamefont {L.~E.}\ \bibnamefont {{Kidder}}}, \bibinfo {author} {\bibfnamefont {H.~P.}\ \bibnamefont {{Pfeiffer}}},\ and\ \bibinfo {author} {\bibfnamefont {M.~A.}\ \bibnamefont {{Scheel}}},\ }\bibfield  {title} {\bibinfo {title} {{High angular momentum hot differentially rotating equilibrium star evolutions in conformally flat spacetime}},\ }\href {https://doi.org/10.1103/PhysRevD.110.043015} {\bibfield  {journal} {\bibinfo  {journal} {\prd}\ }\textbf {\bibinfo {volume} {110}},\ \bibinfo {eid} {043015} (\bibinfo {year} {2024})},\ \Eprint {https://arxiv.org/abs/2402.18529} {arXiv:2402.18529 [astro-ph.HE]} \BibitemShut {NoStop}%
\bibitem [{\citenamefont {{Miravet-Ten{\'e}s}}\ \emph {et~al.}(2022)\citenamefont {{Miravet-Ten{\'e}s}}, \citenamefont {{Cerd{\'a}-Dur{\'a}n}}, \citenamefont {{Obergaulinger}},\ and\ \citenamefont {{Font}}}]{Miravet-Tenes:MRIModel}%
  \BibitemOpen
  \bibfield  {author} {\bibinfo {author} {\bibfnamefont {M.}~\bibnamefont {{Miravet-Ten{\'e}s}}}, \bibinfo {author} {\bibfnamefont {P.}~\bibnamefont {{Cerd{\'a}-Dur{\'a}n}}}, \bibinfo {author} {\bibfnamefont {M.}~\bibnamefont {{Obergaulinger}}},\ and\ \bibinfo {author} {\bibfnamefont {J.~A.}\ \bibnamefont {{Font}}},\ }\bibfield  {title} {\bibinfo {title} {{Assessment of a new sub-grid model for magnetohydrodynamical turbulence. I. Magnetorotational instability}},\ }\href {https://doi.org/10.1093/mnras/stac2888} {\bibfield  {journal} {\bibinfo  {journal} {\mnras}\ }\textbf {\bibinfo {volume} {517}},\ \bibinfo {pages} {3505} (\bibinfo {year} {2022})},\ \Eprint {https://arxiv.org/abs/2210.02173} {arXiv:2210.02173 [astro-ph.HE]} \BibitemShut {NoStop}%
\bibitem [{\citenamefont {{Miravet-Ten{\'e}s}}\ \emph {et~al.}(2024)\citenamefont {{Miravet-Ten{\'e}s}}, \citenamefont {{Cerd{\'a}-Dur{\'a}n}}, \citenamefont {{Obergaulinger}},\ and\ \citenamefont {{Font}}}]{Miravet-Tenes:KHModel}%
  \BibitemOpen
  \bibfield  {author} {\bibinfo {author} {\bibfnamefont {M.}~\bibnamefont {{Miravet-Ten{\'e}s}}}, \bibinfo {author} {\bibfnamefont {P.}~\bibnamefont {{Cerd{\'a}-Dur{\'a}n}}}, \bibinfo {author} {\bibfnamefont {M.}~\bibnamefont {{Obergaulinger}}},\ and\ \bibinfo {author} {\bibfnamefont {J.~A.}\ \bibnamefont {{Font}}},\ }\bibfield  {title} {\bibinfo {title} {{Assessment of a new sub-grid model for magnetohydrodynamical turbulence - II. Kelvin-Helmholtz instability}},\ }\href {https://doi.org/10.1093/mnras/stad3237} {\bibfield  {journal} {\bibinfo  {journal} {\mnras}\ }\textbf {\bibinfo {volume} {527}},\ \bibinfo {pages} {1081} (\bibinfo {year} {2024})},\ \Eprint {https://arxiv.org/abs/2308.06041} {arXiv:2308.06041 [astro-ph.HE]} \BibitemShut {NoStop}%
\bibitem [{\citenamefont {{Miravet-Ten{\'e}s}}\ \emph {et~al.}(2025)\citenamefont {{Miravet-Ten{\'e}s}}, \citenamefont {{Obergaulinger}}, \citenamefont {{Cerd{\'a}-Dur{\'a}n}}, \citenamefont {{Font}},\ and\ \citenamefont {{Ruiz}}}]{Miravet-Tenes:2025}%
  \BibitemOpen
  \bibfield  {author} {\bibinfo {author} {\bibfnamefont {M.}~\bibnamefont {{Miravet-Ten{\'e}s}}}, \bibinfo {author} {\bibfnamefont {M.}~\bibnamefont {{Obergaulinger}}}, \bibinfo {author} {\bibfnamefont {P.}~\bibnamefont {{Cerd{\'a}-Dur{\'a}n}}}, \bibinfo {author} {\bibfnamefont {J.~A.}\ \bibnamefont {{Font}}},\ and\ \bibinfo {author} {\bibfnamefont {M.}~\bibnamefont {{Ruiz}}},\ }\bibfield  {title} {\bibinfo {title} {{Subgrid modelling of MRI-driven turbulence in differentially rotating neutron stars}},\ }\href {https://doi.org/10.48550/arXiv.2509.07081} {\bibfield  {journal} {\bibinfo  {journal} {arXiv e-prints}\ ,\ \bibinfo {eid} {arXiv:2509.07081}} (\bibinfo {year} {2025})},\ \Eprint {https://arxiv.org/abs/2509.07081} {arXiv:2509.07081 [astro-ph.HE]} \BibitemShut {NoStop}%
\bibitem [{\citenamefont {{Shakura}}\ and\ \citenamefont {{Sunyaev}}(1973)}]{ShakuraSunyaev:1973}%
  \BibitemOpen
  \bibfield  {author} {\bibinfo {author} {\bibfnamefont {N.~I.}\ \bibnamefont {{Shakura}}}\ and\ \bibinfo {author} {\bibfnamefont {R.~A.}\ \bibnamefont {{Sunyaev}}},\ }\bibfield  {title} {\bibinfo {title} {{Black holes in binary systems. Observational appearance.}},\ }\href@noop {} {\bibfield  {journal} {\bibinfo  {journal} {\aap}\ }\textbf {\bibinfo {volume} {24}},\ \bibinfo {pages} {337} (\bibinfo {year} {1973})}\BibitemShut {NoStop}%
\bibitem [{\citenamefont {{Balbus}}\ and\ \citenamefont {{Hawley}}(1998)}]{BalbusHawley:1998}%
  \BibitemOpen
  \bibfield  {author} {\bibinfo {author} {\bibfnamefont {S.~A.}\ \bibnamefont {{Balbus}}}\ and\ \bibinfo {author} {\bibfnamefont {J.~F.}\ \bibnamefont {{Hawley}}},\ }\bibfield  {title} {\bibinfo {title} {{Instability, turbulence, and enhanced transport in accretion disks}},\ }\href {https://doi.org/10.1103/RevModPhys.70.1} {\bibfield  {journal} {\bibinfo  {journal} {Reviews of Modern Physics}\ }\textbf {\bibinfo {volume} {70}},\ \bibinfo {pages} {1} (\bibinfo {year} {1998})}\BibitemShut {NoStop}%
\bibitem [{\citenamefont {{Kiuchi}}\ \emph {et~al.}(2018)\citenamefont {{Kiuchi}}, \citenamefont {{Kyutoku}}, \citenamefont {{Sekiguchi}},\ and\ \citenamefont {{Shibata}}}]{Kiuchi:2018}%
  \BibitemOpen
  \bibfield  {author} {\bibinfo {author} {\bibfnamefont {K.}~\bibnamefont {{Kiuchi}}}, \bibinfo {author} {\bibfnamefont {K.}~\bibnamefont {{Kyutoku}}}, \bibinfo {author} {\bibfnamefont {Y.}~\bibnamefont {{Sekiguchi}}},\ and\ \bibinfo {author} {\bibfnamefont {M.}~\bibnamefont {{Shibata}}},\ }\bibfield  {title} {\bibinfo {title} {{Global simulations of strongly magnetized remnant massive neutron stars formed in binary neutron star mergers}},\ }\href {https://doi.org/10.1103/PhysRevD.97.124039} {\bibfield  {journal} {\bibinfo  {journal} {\prd}\ }\textbf {\bibinfo {volume} {97}},\ \bibinfo {eid} {124039} (\bibinfo {year} {2018})},\ \Eprint {https://arxiv.org/abs/1710.01311} {arXiv:1710.01311 [astro-ph.HE]} \BibitemShut {NoStop}%
\bibitem [{\citenamefont {{Radice}}(2020)}]{RadiceMixLength:2020}%
  \BibitemOpen
  \bibfield  {author} {\bibinfo {author} {\bibfnamefont {D.}~\bibnamefont {{Radice}}},\ }\bibfield  {title} {\bibinfo {title} {{Binary Neutron Star Merger Simulations with a Calibrated Turbulence Model}},\ }\href {https://doi.org/10.3390/sym12081249} {\bibfield  {journal} {\bibinfo  {journal} {Symmetry}\ }\textbf {\bibinfo {volume} {12}},\ \bibinfo {pages} {1249} (\bibinfo {year} {2020})},\ \Eprint {https://arxiv.org/abs/2005.09002} {arXiv:2005.09002 [astro-ph.HE]} \BibitemShut {NoStop}%
\bibitem [{\citenamefont {{Radice}}\ and\ \citenamefont {{Bernuzzi}}(2023)}]{RadiceMixLength:2023}%
  \BibitemOpen
  \bibfield  {author} {\bibinfo {author} {\bibfnamefont {D.}~\bibnamefont {{Radice}}}\ and\ \bibinfo {author} {\bibfnamefont {S.}~\bibnamefont {{Bernuzzi}}},\ }\bibfield  {title} {\bibinfo {title} {{Ab-initio General-relativistic Neutrino-radiation Hydrodynamics Simulations of Long-lived Neutron Star Merger Remnants to Neutrino Cooling Timescales}},\ }\href {https://doi.org/10.3847/1538-4357/ad0235} {\bibfield  {journal} {\bibinfo  {journal} {\apj}\ }\textbf {\bibinfo {volume} {959}},\ \bibinfo {eid} {46} (\bibinfo {year} {2023})},\ \Eprint {https://arxiv.org/abs/2306.13709} {arXiv:2306.13709 [astro-ph.HE]} \BibitemShut {NoStop}%
\bibitem [{\citenamefont {{Kiuchi}}\ \emph {et~al.}(2023)\citenamefont {{Kiuchi}}, \citenamefont {{Fujibayashi}}, \citenamefont {{Hayashi}}, \citenamefont {{Kyutoku}}, \citenamefont {{Sekiguchi}},\ and\ \citenamefont {{Shibata}}}]{Kiuchi:2023}%
  \BibitemOpen
  \bibfield  {author} {\bibinfo {author} {\bibfnamefont {K.}~\bibnamefont {{Kiuchi}}}, \bibinfo {author} {\bibfnamefont {S.}~\bibnamefont {{Fujibayashi}}}, \bibinfo {author} {\bibfnamefont {K.}~\bibnamefont {{Hayashi}}}, \bibinfo {author} {\bibfnamefont {K.}~\bibnamefont {{Kyutoku}}}, \bibinfo {author} {\bibfnamefont {Y.}~\bibnamefont {{Sekiguchi}}},\ and\ \bibinfo {author} {\bibfnamefont {M.}~\bibnamefont {{Shibata}}},\ }\bibfield  {title} {\bibinfo {title} {{Self-Consistent Picture of the Mass Ejection from a One Second Long Binary Neutron Star Merger Leaving a Short-Lived Remnant in a General-Relativistic Neutrino-Radiation Magnetohydrodynamic Simulation}},\ }\href {https://doi.org/10.1103/PhysRevLett.131.011401} {\bibfield  {journal} {\bibinfo  {journal} {\prl}\ }\textbf {\bibinfo {volume} {131}},\ \bibinfo {eid} {011401} (\bibinfo {year} {2023})},\ \Eprint {https://arxiv.org/abs/2211.07637} {arXiv:2211.07637 [astro-ph.HE]} \BibitemShut {NoStop}%
\bibitem [{\citenamefont {Spruit}(2002)}]{Spruit:2001tz}%
  \BibitemOpen
  \bibfield  {author} {\bibinfo {author} {\bibfnamefont {H.~C.}\ \bibnamefont {Spruit}},\ }\bibfield  {title} {\bibinfo {title} {{Dynamo action by differential rotation in a stably stratified stellar interior}},\ }\href {https://doi.org/10.1051/0004-6361:20011465} {\bibfield  {journal} {\bibinfo  {journal} {Astron. Astrophys.}\ }\textbf {\bibinfo {volume} {381}},\ \bibinfo {pages} {923} (\bibinfo {year} {2002})},\ \Eprint {https://arxiv.org/abs/astro-ph/0108207} {arXiv:astro-ph/0108207} \BibitemShut {NoStop}%
\bibitem [{\citenamefont {{Wheeler}}\ \emph {et~al.}(2015)\citenamefont {{Wheeler}}, \citenamefont {{Kagan}},\ and\ \citenamefont {{Chatzopoulos}}}]{Wheeler:2015}%
  \BibitemOpen
  \bibfield  {author} {\bibinfo {author} {\bibfnamefont {J.~C.}\ \bibnamefont {{Wheeler}}}, \bibinfo {author} {\bibfnamefont {D.}~\bibnamefont {{Kagan}}},\ and\ \bibinfo {author} {\bibfnamefont {E.}~\bibnamefont {{Chatzopoulos}}},\ }\bibfield  {title} {\bibinfo {title} {{The Role of the Magnetorotational Instability in Massive Stars}},\ }\href {https://doi.org/10.1088/0004-637X/799/1/85} {\bibfield  {journal} {\bibinfo  {journal} {\apj}\ }\textbf {\bibinfo {volume} {799}},\ \bibinfo {eid} {85} (\bibinfo {year} {2015})},\ \Eprint {https://arxiv.org/abs/1411.5714} {arXiv:1411.5714 [astro-ph.SR]} \BibitemShut {NoStop}%
\bibitem [{\citenamefont {{Margalit}}\ \emph {et~al.}(2022)\citenamefont {{Margalit}}, \citenamefont {{Jermyn}}, \citenamefont {{Metzger}}, \citenamefont {{Roberts}},\ and\ \citenamefont {{Quataert}}}]{Margalit:2022}%
  \BibitemOpen
  \bibfield  {author} {\bibinfo {author} {\bibfnamefont {B.}~\bibnamefont {{Margalit}}}, \bibinfo {author} {\bibfnamefont {A.~S.}\ \bibnamefont {{Jermyn}}}, \bibinfo {author} {\bibfnamefont {B.~D.}\ \bibnamefont {{Metzger}}}, \bibinfo {author} {\bibfnamefont {L.~F.}\ \bibnamefont {{Roberts}}},\ and\ \bibinfo {author} {\bibfnamefont {E.}~\bibnamefont {{Quataert}}},\ }\bibfield  {title} {\bibinfo {title} {{Angular-momentum Transport in Proto-neutron Stars and the Fate of Neutron Star Merger Remnants}},\ }\href {https://doi.org/10.3847/1538-4357/ac8b01} {\bibfield  {journal} {\bibinfo  {journal} {\apj}\ }\textbf {\bibinfo {volume} {939}},\ \bibinfo {eid} {51} (\bibinfo {year} {2022})},\ \Eprint {https://arxiv.org/abs/2206.10645} {arXiv:2206.10645 [astro-ph.HE]} \BibitemShut {NoStop}%
\bibitem [{\citenamefont {{Aguilera-Miret}}\ \emph {et~al.}(2024)\citenamefont {{Aguilera-Miret}}, \citenamefont {{Palenzuela}}, \citenamefont {{Carrasco}}, \citenamefont {{Rosswog}},\ and\ \citenamefont {{Vigan{\`o}}}}]{Aguilera-Miret:2024}%
  \BibitemOpen
  \bibfield  {author} {\bibinfo {author} {\bibfnamefont {R.}~\bibnamefont {{Aguilera-Miret}}}, \bibinfo {author} {\bibfnamefont {C.}~\bibnamefont {{Palenzuela}}}, \bibinfo {author} {\bibfnamefont {F.}~\bibnamefont {{Carrasco}}}, \bibinfo {author} {\bibfnamefont {S.}~\bibnamefont {{Rosswog}}},\ and\ \bibinfo {author} {\bibfnamefont {D.}~\bibnamefont {{Vigan{\`o}}}},\ }\bibfield  {title} {\bibinfo {title} {{Delayed jet launching in binary neutron star mergers with realistic initial magnetic fields}},\ }\href {https://doi.org/10.1103/PhysRevD.110.083014} {\bibfield  {journal} {\bibinfo  {journal} {\prd}\ }\textbf {\bibinfo {volume} {110}},\ \bibinfo {eid} {083014} (\bibinfo {year} {2024})},\ \Eprint {https://arxiv.org/abs/2407.20335} {arXiv:2407.20335 [astro-ph.HE]} \BibitemShut {NoStop}%
\bibitem [{\citenamefont {{Metzger}}\ \emph {et~al.}(2009)\citenamefont {{Metzger}}, \citenamefont {{Piro}},\ and\ \citenamefont {{Quataert}}}]{Metzger:2009}%
  \BibitemOpen
  \bibfield  {author} {\bibinfo {author} {\bibfnamefont {B.~D.}\ \bibnamefont {{Metzger}}}, \bibinfo {author} {\bibfnamefont {A.~L.}\ \bibnamefont {{Piro}}},\ and\ \bibinfo {author} {\bibfnamefont {E.}~\bibnamefont {{Quataert}}},\ }\bibfield  {title} {\bibinfo {title} {{Neutron-rich freeze-out in viscously spreading accretion discs formed from compact object mergers}},\ }\href {https://doi.org/10.1111/j.1365-2966.2008.14380.x} {\bibfield  {journal} {\bibinfo  {journal} {\mnras}\ }\textbf {\bibinfo {volume} {396}},\ \bibinfo {pages} {304} (\bibinfo {year} {2009})},\ \Eprint {https://arxiv.org/abs/0810.2535} {arXiv:0810.2535 [astro-ph]} \BibitemShut {NoStop}%
\bibitem [{\citenamefont {{Griffiths}}\ \emph {et~al.}(2022)\citenamefont {{Griffiths}}, \citenamefont {{Eggenberger}}, \citenamefont {{Meynet}}, \citenamefont {{Moyano}},\ and\ \citenamefont {{Aloy}}}]{Griffiths:2022}%
  \BibitemOpen
  \bibfield  {author} {\bibinfo {author} {\bibfnamefont {A.}~\bibnamefont {{Griffiths}}}, \bibinfo {author} {\bibfnamefont {P.}~\bibnamefont {{Eggenberger}}}, \bibinfo {author} {\bibfnamefont {G.}~\bibnamefont {{Meynet}}}, \bibinfo {author} {\bibfnamefont {F.}~\bibnamefont {{Moyano}}},\ and\ \bibinfo {author} {\bibfnamefont {M.-{\'A}.}\ \bibnamefont {{Aloy}}},\ }\bibfield  {title} {\bibinfo {title} {{The magneto-rotational instability in massive stars}},\ }\href {https://doi.org/10.1051/0004-6361/202243599} {\bibfield  {journal} {\bibinfo  {journal} {\aap}\ }\textbf {\bibinfo {volume} {665}},\ \bibinfo {eid} {A147} (\bibinfo {year} {2022})},\ \Eprint {https://arxiv.org/abs/2204.00016} {arXiv:2204.00016 [astro-ph.SR]} \BibitemShut {NoStop}%
\bibitem [{\citenamefont {{Stone}}\ \emph {et~al.}(1996)\citenamefont {{Stone}}, \citenamefont {{Hawley}}, \citenamefont {{Gammie}},\ and\ \citenamefont {{Balbus}}}]{Stone:1996}%
  \BibitemOpen
  \bibfield  {author} {\bibinfo {author} {\bibfnamefont {J.~M.}\ \bibnamefont {{Stone}}}, \bibinfo {author} {\bibfnamefont {J.~F.}\ \bibnamefont {{Hawley}}}, \bibinfo {author} {\bibfnamefont {C.~F.}\ \bibnamefont {{Gammie}}},\ and\ \bibinfo {author} {\bibfnamefont {S.~A.}\ \bibnamefont {{Balbus}}},\ }\bibfield  {title} {\bibinfo {title} {{Three-dimensional Magnetohydrodynamical Simulations of Vertically Stratified Accretion Disks}},\ }\href {https://doi.org/10.1086/177280} {\bibfield  {journal} {\bibinfo  {journal} {\apj}\ }\textbf {\bibinfo {volume} {463}},\ \bibinfo {pages} {656} (\bibinfo {year} {1996})}\BibitemShut {NoStop}%
\bibitem [{\citenamefont {{Abramowicz}}\ \emph {et~al.}(1996)\citenamefont {{Abramowicz}}, \citenamefont {{Brandenburg}},\ and\ \citenamefont {{Lasota}}}]{Abramowicz:1996}%
  \BibitemOpen
  \bibfield  {author} {\bibinfo {author} {\bibfnamefont {M.}~\bibnamefont {{Abramowicz}}}, \bibinfo {author} {\bibfnamefont {A.}~\bibnamefont {{Brandenburg}}},\ and\ \bibinfo {author} {\bibfnamefont {J.-P.}\ \bibnamefont {{Lasota}}},\ }\bibfield  {title} {\bibinfo {title} {{The dependence of the viscosity in accretion discs on the shear/vorticity ratio}},\ }\href {https://doi.org/10.1093/mnras/281.2.L21} {\bibfield  {journal} {\bibinfo  {journal} {\mnras}\ }\textbf {\bibinfo {volume} {281}},\ \bibinfo {pages} {L21} (\bibinfo {year} {1996})}\BibitemShut {NoStop}%
\bibitem [{\citenamefont {{Masada}}\ \emph {et~al.}(2012)\citenamefont {{Masada}}, \citenamefont {{Takiwaki}}, \citenamefont {{Kotake}},\ and\ \citenamefont {{Sano}}}]{Masada:2012}%
  \BibitemOpen
  \bibfield  {author} {\bibinfo {author} {\bibfnamefont {Y.}~\bibnamefont {{Masada}}}, \bibinfo {author} {\bibfnamefont {T.}~\bibnamefont {{Takiwaki}}}, \bibinfo {author} {\bibfnamefont {K.}~\bibnamefont {{Kotake}}},\ and\ \bibinfo {author} {\bibfnamefont {T.}~\bibnamefont {{Sano}}},\ }\bibfield  {title} {\bibinfo {title} {{Local Simulations of the Magnetorotational Instability in Core-collapse Supernovae}},\ }\href {https://doi.org/10.1088/0004-637X/759/2/110} {\bibfield  {journal} {\bibinfo  {journal} {\apj}\ }\textbf {\bibinfo {volume} {759}},\ \bibinfo {eid} {110} (\bibinfo {year} {2012})},\ \Eprint {https://arxiv.org/abs/1209.2360} {arXiv:1209.2360 [astro-ph.SR]} \BibitemShut {NoStop}%
\bibitem [{\citenamefont {Barr\`ere}\ \emph {et~al.}(2022)\citenamefont {Barr\`ere}, \citenamefont {Guilet}, \citenamefont {Reboul-Salze}, \citenamefont {Raynaud},\ and\ \citenamefont {Janka}}]{Barrere:2022gwv}%
  \BibitemOpen
  \bibfield  {author} {\bibinfo {author} {\bibfnamefont {P.}~\bibnamefont {Barr\`ere}}, \bibinfo {author} {\bibfnamefont {J.}~\bibnamefont {Guilet}}, \bibinfo {author} {\bibfnamefont {A.}~\bibnamefont {Reboul-Salze}}, \bibinfo {author} {\bibfnamefont {R.}~\bibnamefont {Raynaud}},\ and\ \bibinfo {author} {\bibfnamefont {H.~T.}\ \bibnamefont {Janka}},\ }\bibfield  {title} {\bibinfo {title} {{A new scenario for magnetar formation: Tayler-Spruit dynamo in a proto-neutron star spun up by fallback}},\ }\href {https://doi.org/10.1051/0004-6361/202244172} {\bibfield  {journal} {\bibinfo  {journal} {Astron. Astrophys.}\ }\textbf {\bibinfo {volume} {668}},\ \bibinfo {pages} {A79} (\bibinfo {year} {2022})},\ \Eprint {https://arxiv.org/abs/2206.01269} {arXiv:2206.01269 [astro-ph.HE]} \BibitemShut {NoStop}%
\bibitem [{\citenamefont {{Reboul-Salze}}\ \emph {et~al.}(2025)\citenamefont {{Reboul-Salze}}, \citenamefont {{Barr{\`e}re}}, \citenamefont {{Kiuchi}}, \citenamefont {{Guilet}}, \citenamefont {{Raynaud}}, \citenamefont {{Fujibayashi}},\ and\ \citenamefont {{Shibata}}}]{Reboul-Salze:2024jst}%
  \BibitemOpen
  \bibfield  {author} {\bibinfo {author} {\bibfnamefont {A.}~\bibnamefont {{Reboul-Salze}}}, \bibinfo {author} {\bibfnamefont {P.}~\bibnamefont {{Barr{\`e}re}}}, \bibinfo {author} {\bibfnamefont {K.}~\bibnamefont {{Kiuchi}}}, \bibinfo {author} {\bibfnamefont {J.}~\bibnamefont {{Guilet}}}, \bibinfo {author} {\bibfnamefont {R.}~\bibnamefont {{Raynaud}}}, \bibinfo {author} {\bibfnamefont {S.}~\bibnamefont {{Fujibayashi}}},\ and\ \bibinfo {author} {\bibfnamefont {M.}~\bibnamefont {{Shibata}}},\ }\bibfield  {title} {\bibinfo {title} {{Tayler-Spruit dynamo in binary neutron star merger remnants}},\ }\href {https://doi.org/10.1051/0004-6361/202453126} {\bibfield  {journal} {\bibinfo  {journal} {\aap}\ }\textbf {\bibinfo {volume} {699}},\ \bibinfo {eid} {A4} (\bibinfo {year} {2025})},\ \Eprint {https://arxiv.org/abs/2411.19328} {arXiv:2411.19328 [astro-ph.HE]} \BibitemShut {NoStop}%
\bibitem [{\citenamefont {Barr\`ere}\ \emph {et~al.}(2025)\citenamefont {Barr\`ere}, \citenamefont {Guilet}, \citenamefont {Raynaud},\ and\ \citenamefont {Reboul-Salze}}]{Barrere:2024jyw}%
  \BibitemOpen
  \bibfield  {author} {\bibinfo {author} {\bibfnamefont {P.}~\bibnamefont {Barr\`ere}}, \bibinfo {author} {\bibfnamefont {J.}~\bibnamefont {Guilet}}, \bibinfo {author} {\bibfnamefont {R.}~\bibnamefont {Raynaud}},\ and\ \bibinfo {author} {\bibfnamefont {A.}~\bibnamefont {Reboul-Salze}},\ }\bibfield  {title} {\bibinfo {title} {{Tayler-Spruit dynamo in stably stratified rotating fluids: Application to proto-magnetars}},\ }\href {https://doi.org/10.1051/0004-6361/202451337} {\bibfield  {journal} {\bibinfo  {journal} {Astron. Astrophys.}\ }\textbf {\bibinfo {volume} {695}},\ \bibinfo {pages} {A183} (\bibinfo {year} {2025})},\ \Eprint {https://arxiv.org/abs/2407.01775} {arXiv:2407.01775 [astro-ph.HE]} \BibitemShut {NoStop}%
\bibitem [{\citenamefont {Fuller}\ \emph {et~al.}(2019)\citenamefont {Fuller}, \citenamefont {Piro},\ and\ \citenamefont {Jermyn}}]{Fuller:2019ckz}%
  \BibitemOpen
  \bibfield  {author} {\bibinfo {author} {\bibfnamefont {J.}~\bibnamefont {Fuller}}, \bibinfo {author} {\bibfnamefont {A.~L.}\ \bibnamefont {Piro}},\ and\ \bibinfo {author} {\bibfnamefont {A.~S.}\ \bibnamefont {Jermyn}},\ }\bibfield  {title} {\bibinfo {title} {{Slowing the spins of stellar cores}},\ }\href {https://doi.org/10.1093/mnras/stz514} {\bibfield  {journal} {\bibinfo  {journal} {\mnras}\ }\textbf {\bibinfo {volume} {485}},\ \bibinfo {pages} {3661} (\bibinfo {year} {2019})},\ \Eprint {https://arxiv.org/abs/1902.08227} {arXiv:1902.08227} \BibitemShut {NoStop}%
\end{thebibliography}%

\end{document}